\documentclass{sigchi}

\usepackage{balance}  
\usepackage{graphics} 
\usepackage{times}    
\usepackage{url}      
\usepackage{subfigure}
\usepackage{graphicx}
\usepackage{multirow}
\usepackage{url}
\usepackage{xcolor}

\usepackage{cite}
\usepackage{tabu}
\usepackage{booktabs}
\usepackage{multirow}

\def\plainkeywords{secure web protocols; wearable devices; energy utilization; performance evaluation}
\author{}

\makeatletter
\def\url@leostyle{%
  \@ifundefined{selectfont}{\def\UrlFont{\sf}}{\def\UrlFont{\small\bf\ttfamily}}}
\makeatother
\def\pprw{8.5in}
\def\pprh{11in}

\usepackage[pdftex]{hyperref}
\hypersetup{
pdftitle={\textit{Are wearable devices ready for HTTPS?} \\ Measuring the cost of secure communication protocols on wearable devices},
pdfauthor={LaTeX},
pdfkeywords={SIGCHI, proceedings, archival format},
bookmarksnumbered,
pdfstartview={FitH},
colorlinks,
citecolor=black,
filecolor=black,
linkcolor=black,
urlcolor=black,
breaklinks=true,
}

\newcommand*{\affaddrarray}[1]{#1} 
\newcommand*{\affmark}[1][*]{\textsuperscript{#1}}

\begin{document}

\title{Are wearable devices ready for HTTPS? \\ Measuring the cost of secure communication protocols on wearable devices}

\author{
\alignauthor{Harini Kolamunna,\affmark[1]~Jagmohan Chauhan,\affmark[1]~Yining Hu,~Kanchana Thilakarathna,\affmark[1]~Diego Perino,\affmark[2]$^{,}$\affmark[4]\thanks{This work was done while the author was at Data61.} 
\\Dwight Makaroff~\affmark[3]$^{,}$\affmark[5]\footnotemark[1]~ and Aruna Seneviratne\affmark[1] \\
\affaddr{Data61/CSIRO-Australia} \\
\affaddrarray{\affmark[1]University of New South Wales},
\affaddrarray{\affmark[2]Telefonica Research},
\affaddrarray{\affmark[3]University of Saskatchewan}\\
Email: \email{(firstname.lastname)@data61.csiro.au}, 
\affaddrarray{\affmark[4]diego.perino@telefonica.com},
\affaddrarray{\affmark[5]makaroff@cs.usask.ca}
}
}

\maketitle

\begin{abstract}
The majority of available wearable devices require communication with Internet servers for data analysis and storage, and rely on a paired smartphone to enable secure communication. However, wearable devices are mostly equipped with WiFi network interfaces, enabling direct communication with the Internet. Secure communication protocols should then run on these wearables itself, yet it is not clear if they can be efficiently supported. 

In this paper, we show that wearable devices are ready for direct and secure Internet communication by means of experiments with both controlled and Internet servers. We observe that the overall energy consumption and communication delay can be reduced with direct Internet connection via WiFi from wearables compared to using smartphones as relays via Bluetooth. We also show that the additional HTTPS cost caused by TLS handshake and encryption is closely related to number of parallel connections, and has the same relative impact on wearables and smartphones.

\end{abstract}

\keywords{\plainkeywords}

\section{Introduction}
Wearable devices such as smartwatches, smartglasses and fitness bands, are becoming increasingly popular and were predicted in 2014 to become commodity devices in the same manner as smartphones~\cite{wearmarket}. 
These devices are equipped with a rich set of sensors that can continuously monitor a wide variety of attributes of the human body, the physical surroundings and online user behaviours not available through any other means \cite{sazonov2014wearable}.

Many wearable device applications (apps) upload users' personal information collected from body-worn sensors to cloud servers for analysis. Due to the sensitivity of the personal data collected by wearables, they must be transferred using a secure communication protocol. Today, most of the apps rely on a smartphone counterpart for communication with cloud services and in this way take advantage of secure communication protocols such as HTTPS. 
A wearable device typically communicates with the paired smartphone via Bluetooth. Such communication is usually considered secure due to the low transmission range.
However, the majority of wearable devices are already equipped with WiFi capability while Cellular networking is becoming a reality. These technologies enable direct Internet connectivity from wearables; it is thus vital to understand whether or not secure communication protocols can be efficiently supported by wearables directly.  

In this paper, we address this critical question through an experimental study. As nearly 50\% of today's Internet traffic is HTTPS \cite{Naylor}, it is a natural candidate for secure communication for standalone wearable devices. Naylor  \textit{et al.} show that the cost of HTTPS is not negligible even in the case of smartphones \cite{Naylor}. We thus study the impact of HTTPS on wearables 
compared to HTTP in terms of the amount of \textit{downloaded data}, \textit{data transfer time} and \textit{energy consumption}. 
We leverage two popular categories of wearables (i.e., smartglass and smartwatch), and take measurements from both a controlled web server and landing webpages of popular Internet websites. This allows us to characterize the resource consumption for each sub-phase of the protocols precisely to derive the main factors contributing to the HTTPS performance difference, and to measure the impact
when multiple parallel TCP 
connections and multiple external servers are used to access web resources from these devices.

We make the following main observations. First, the relative cost of HTTPS to HTTP in terms of \textit{downloaded data}, \textit{data transfer time} on wearables is comparable to that of smartphone, validating the fact that wearables are ready for HTTPS in terms of computing and networking capabilities. Second, because of the smaller battery capacity on wearables, the normalized  \textit{energy consumption} by the 100\% battery capacity is higher in wearables than the smartphone. However, the overall \textit{energy consumption} considering both wearables and smartphone is reduced (by $\sim$78\% in smartwatch and 100\% in smartphone) with direct and secure Internet communication from wearables primarily due to the elimination of the overhead of data exchange between smartwatch and smartphone.
Moreover, relaying incurs additional delay in real-time communication with external servers compared to direct connectivity from wearables. Finally, we verify that the additional cost of HTTPS is mostly  due to the TLS handshake (KEY exchange) phase and that the magnitude of the cost is closely related to the amount of data exchanged, and the number of parallel connections.
These observations lead us to conclude that \emph{wearable devices of today are ready for direct and secure internet communication via WiFi}. 

The remainder of the paper is organized as follows. We first overview the background and related work. The experimental methodology is described next. We then present our analysis of HTTPS overhead in a controlled testbed environment followed by the results of web-browsing experiments.
Finally, we synthesize the implications of these findings into a set of recommendations with suggestions for future work.

\section{Background and Related Work}

\label{sec:background}

\subsection{Transport Layer Security (TLS) Background}
The most widely used protocol to achieve secure communications over the Internet is HTTPS (HTTP over TLS)\cite{rfctls}.  We first describe the phases in HTTPS and HTTP connections followed by TLS session negotiation procedure and the associated cryptographic algorithms. 

\begin{figure}[t]
\centering  \vspace{-4mm}
\includegraphics[width=0.48\textwidth]{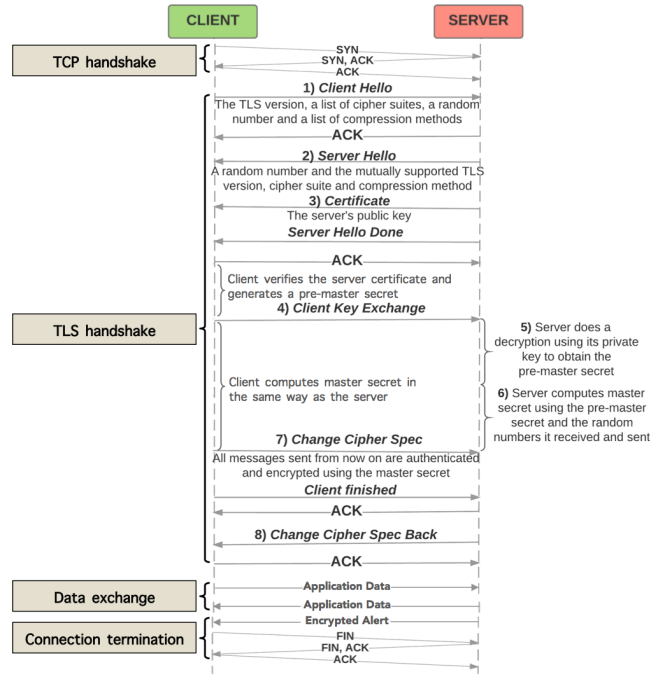}
\caption{HTTPS message sequence diagram with detailed TLS handshaking steps.}\vspace{-5mm}
\label{fig:tls_handshake} 
\end{figure}

Consider establishing a HTTPS connection 
(see Figure \ref{fig:tls_handshake}). The first phase is the \textit{TCP handshake}, which is a three way handshake between the server and the client. During the next phase, which is the \textit{TLS handshake}, server authentication and encryption key exchange takes place. 
Then the encrypted messages are exchanged during the \textit{Data exchange} phase. Once message passing is finished, the TLS and TCP connections terminate in the \textit{Connection termination} phase. An HTTP connection only has three phases: \textit{TCP handshake}, followed by \textit{Data exchange} (which exchanges plaintext messages), and \textit{Connection termination}.

Next we describe the \textit{TLS handshake} sub-phase in detail.

\begin{enumerate}
\item A TLS session is initiated by the client by sending a \textit{Client Hello} message to the server. The \textit{Client Hello} message specifies the client's TLS capabilities, including the TLS version, cipher suites, a random number that will be used later to compute the final symmetric key, and compression method options. A cipher suite encompasses the choices/options for the cryptographic algorithms that are to be used during different phases of TLS session negotiation and data transfer:  the  \textit{key exchange algorithm}, the \textit{authentication algorithm}, the \textit{bulk cipher algorithm}, and the \textit{MAC (message authentication code) algorithm}.  
\vspace{-0.5mm}

\item  Then, the server replies with a \textit{Server Hello} message that contains the TLS  version,  cipher suite and compression method selected by the server from the choices offered by the client that are also supported by the server. Additionally, the server sends  a random number in the message.
\vspace{-0.5mm}

\item The server then sends a message contains \textit{Certificate}, \textit{Server Key Exchange} that includes server's public key, and also sends a \textit{Server Hello Done} message to indicate to the client that the server has finished the Hello process.
\vspace{-0.5mm}

\item The client first verifies the server certificate. It then generates a pre-master secret that is encrypted using the server public key. The pre-master secret is then sent over to the server using a \textit{Client Key Exchange} message. 
\vspace{-0.5mm}

\item  The server decrypts the \textit{Client Key Exchange} message using its private key to obtain the pre-master secret.
 \vspace{-0.5mm}

\item The server uses a combination of the pre-master secret, the random number it sent to the client in \textit{Server Hello} 
message and the random number it got from the client in \textit{Client Hello} message to compute the master secret€. The client on its device computes the mater secret in the same manner. 
\vspace{-0.5mm}

\item   The client sends a \textit{Change Cipher Spec} notification to the server indicating that all subsequent messages
 will be authenticated and encrypted using the master secret. The client also sends an encrypted \textit{Client Finished} message to the server. 
\vspace{-0.5mm}

\item Finally, the server sends a \textit{Change Cipher Spec} back to the client, completing the handshake. After the TLS negotiation phase is over, the encrypted application data is transferred between the client and the server. 
\end{enumerate}

\noindent\textbf{TLS Cryptographic Algorithms.}
Cryptographic Algorithms are comprised of four types of algorithms:

\noindent{\textit{Key Exchange Algorithm:}} The key exchange algorithm defines how the keys should be generated and exchanged  between the client and server  during the TLS session setup.  The commonly available Key Exchange algorithms include RSA (Rivest-Shamir-Adleman), DH (Diffie-Hellman),  ECDH (Elliptic Curve Diffie-Hellman) and ECDHE (Elliptic Curve Diffie-Hellman Ephemeral). Both RSA and DH (as well as variations of DH) are asymmetric algorithm, which means that they use two different keys: a public key for encryption and a private key for decryption. The advantage of Elliptic curve based algorithms over RSA  is that they provide same security level as RSA with smaller key sizes.  ECDHE provides PFS (perfect forward secrecy). PFS ensures that the sessions recorded in the past cannot be retrieved even if the server's long term private keys are compromised. 

\noindent{\textit{Authentication Algorithm:}} The authentication algorithm determines if the client is communicating with the correct server entity.  The authentication is generally done by exchanging certificates. RSA, Digital Signature Algorithm (DSA) and Elliptic Curve Digital Signature Algorithm (ECDSA) are the commonly used authentication algorithms.  All the authentication algorithms are asymmetric and differ in the mathematical operations underlying the algorithm. 

\noindent{\textit{Bulk Cipher Algorithm:}} After the initial session setup phase, the  user data between the  client and the server is encrypted and decrypted  using bulk cipher algorithms.  All the bulk cipher algorithms are symmetric algorithms  using the same key for encryption and decryption. The  bulk cipher algorithms include block-based  algorithms such as DES (Data Encryption Standard), 3DES, AES (Advanced Encryption Standard) and stream-based algorithms such as RC4 (Rivest's Cipher). 

\noindent{\textit{MAC Algorithm:}} The integrity of the application data is maintained using MAC algorithm. It provides the means for a receiver to know that the application data is not altered en-route. The common MAC algorithms are MD5 (Message Digest 5) and  SHA (Secure Hash Algorithm).

\subsection{Related Work}
Much previous research has investigated the performance of HTTPS  on different computing platforms such as web servers \cite{Apostolopoulos, Coarfa} and mobile devices \cite{Youngsang, Miranda, Potlapally, Naylor}. Apostolopoulos et al. \cite{Apostolopoulos} measured the performance of TLS using SPECWeb96 benchmark on Apache and Netscape web servers. Their results showed that the use of TLS decreases the number of web transactions that a server can handle by a factor of two. A comprehensive study  by  Coarfa \textit{et al.} \cite{Coarfa} on TLS web servers revealed that  RSA operations incur the largest performance cost. The study also found out that RSA accelerators are effective  for e-commerce type workloads as they experience low TLS session reuse.  

Youngsang \textit{et al.} \cite{Youngsang} compared the performance of the TLS protocol between PDA and laptop in secured WEP and open WiFi environments. The studies confirmed  that the  additional delay for TLS handshake is three times more on a PDA than for a laptop. However, the major factor is not cryptographic computation but network latency and other PDA architecture issues.  
In other work, Naylor \textit{et al.} \cite{Naylor} studied the collected dataset from large ISPs to examine the world-wide adoption rate of HTTPS.  The study also indicated that HTTPS adds significant latency in mobile networks and that the cost of TLS on smartphones is not negligible. 

Miranda \textit{et al.} \cite{Miranda} measured the  energy consumption of TLS on Nokia N95 mobile phone  on WiFi and 3G networks and showed that the energy required for transmission and I/O during TLS handshake far exceeds the energy spent on computing cryptographic operations. 
Potlapally \textit{et al.} \cite{Potlapally} performed a very comprehensive performance analysis by studying the impact of various TLS cryptographic algorithms 
with varying bit sizes  on the energy consumption of a PDA device. The main findings of the study suggested  that asymmetric and hash algorithms have the highest and the least energy costs, respectively, and the trade-off between using cryptographic algorithm and energy consumption can be achieved by tuning parameters such as key size.

Our study differs from previous work as we are expanding the context of evaluating web security protocol overhead to the wearable device platform.
To the best of our knowledge, we are the first to quantify the performance of HTTPS versus HTTP experimentally in terms  of \textit{energy consumption}, \textit{data transfer time} and \textit{downloaded data} on wearable devices and compare them to HTTPS performance on smartphones.

\vspace{-2mm}   
   
\section{Experimental methodology}
\label{sec:methodology}
In this section, we first present the devices used in experiments, followed by the scenarios and the evaluation metrics. 

\begin{table}[tb]
\scriptsize
\centering \vspace{-2mm}
\begin{tabu}{cccc}
\tabucline[0.8pt]{-}
\noalign{\smallskip}
\textbf{Device}&\textbf{CPU}&\textbf{Memory}&\textbf{Battery}\\
\vspace{-2mm}
\\\hline\hline\noalign{\smallskip}
\multirow{1}{*}{Watch}&Quad-core processor 1.2 GHz&512 MB&410 mAh\\
\multirow{1}{*}{Glass}&Dual-core processor 1 GHz&2 GB&570 mAh\\
\multirow{1}{*}{Phone}&Quad-core processor 1.5 GHz&2 GB&2100 mAh\\
\noalign{\smallskip}
\tabucline[0.8pt]{-}
\end{tabu}
\caption{ Hardware characteristics of the devices.}\vspace{-2mm}
\label{table:devices}
\vspace{-4mm}
\end{table}

\vspace{-3mm}   

\subsection{Devices} We select two devices representative of two popular wearable categories: smartglasses and smartwatches. Specifically, we use Google Glass Explorer edition running Android XE 18.11 as an example of smartglasses, and  LG G Watch R running Android Wear 5 as an example of smartwatches. 
Baseline results are obtained from a Nexus 4 smartphone running Android 5. All communication to the internet is done via WiFi while smartphone and smartwatch communicate via Bluetooth.
The hardware specifications for all the devices is shown in Table~\ref{table:devices}.

\subsection{Experimental Scenarios}
We perform two sets of experiments; 1) within a controlled testbed of a customized web server and client-side application, and 2) with publicly available Internet web servers and existing client applications. We focus on file download only (i.e., GET), as the same observations and trends also hold for a file upload (i.e., POST). 

{\bf Controlled Testbed.} We run a web server in the local network that supports HTTP and HTTPS file requests. The server runs Apache server 2.4.7 on Ubuntu 14.04; we disable caching and page encoding by setting the specific headings in downloaded PHP files. On the client side, we develop a custom mobile/wearable app that is compatible with all three devices, and that can generate a single GET or single POST request to the web server. We were not able to leverage existing wearable apps as they do not allow the user to select between HTTP and HTTPS protocols, nor to collect performance stats. 
We run the following two sets of experiments in the controlled testbed. Each experiment is repeated 30 times and results averaged over all runs.

1) \textsf{Exp1:} downloading a single file (GET request) from the local web server, which allows us to characterize the cost associated with different phases of a HTTP/HTTPS session precisely. We also vary the file size to understand the main factors contributing to this cost on wearables. For this experiment, as the reference TLS parameters, we use TLS 1.0 version with the following set of algorithms: ECDHE for key exchange, RSA for authentication, AES 128 for bulk cipher, and SHA for MAC. 

2) \textsf{Exp2:} downloading a single file from the local web server with different combinations of TLS cryptographic algorithms. We vary the supported TLS algorithm combinations one by one and tested it on a fixed file size of 100 KB. As there are many options available for each of the four major types of algorithms involved in TLS, we focus on the ones frequently used on modern devices. 
For the key exchange Algorithm, we select RSA (2048 bits) and ECDHE (65 bytes). The former is the oldest and widely used algorithm for key exchange. The latter is 
the preferred choice on modern devices and web servers.  
For the authentication algorithm we select RSA; indeed, RSA signed certificates are the most used on the web. 
We do not consider ECDSA as the usage of ECDSA signed certificates on the web is still in a very nascent stage. We choose to experiment with AES (128 and 256 bit) and RC4 for bulk cipher algorithm.  We test RC4 to compare stream-based algorithms against block-based algorithms such as AES. Finally, for the MAC algorithm, we choose SHA. Indeed, modern web servers use AES as bulk cipher algorithm and SHA as MAC algorithm because they are the most secure algorithms available in their respective categories.

\begin{table}[tb]
	\centering
	\scriptsize
	\begin{tabu}{ccc}
	\tabucline[0.8pt]{-}
	\noalign{\smallskip}
	\textbf{Website} & \textbf{Url} & \textbf{ID} \\
	\hline\hline\noalign{\smallskip}
	Bing & www.bing.com & B \\
	Apple & www.apple.com & A\\
	VK  & www.vk.com & V \\
	Terraclicks & www.terraclicks.com & T \\
	Java & www.java.com & J \\
	Douban & www.douban.com & D \\

	\noalign{\smallskip}
	\tabucline[0.8pt]{-}
	\end{tabu}  
	\caption{Websites used in our experiments.}
	
	\label{tab:websites}
	 
\vspace{-5 mm}
\end{table}

{\bf Internet Experiments.} We investigate the impact of HTTPS on wearable devices in realistic settings with public web servers and existing web browsers.
Specifically, \textsf{Exp3}  allow us to measure the impact of TLS when multiple parallel TCP connections and multiple external servers are used to access web resources from wearables. Each experiment is repeated 5 times and results averaged over all runs. 

\textsf{3) Exp3:} We select a subset of websites from Alexa top 500\footnote{\url{http://www.alexa.com/topsites}} and download the landing webpages of those websites. The selected websites support both HTTP and HTTPS and transfer all objects using one of the two protocols. It turns out that only 6 websites satisfy our requirements (Table~\ref{tab:websites}), while the others either support only one of the two protocols, or use a mix of HTTP and HTTPS for the data exchange. As our goal is to understand the cost of HTTPS with multiple connections and servers and not to provide an in-depth analysis of the impact of HTTPS on user web experience, we believe these websites are sufficient for \textsf{Exp3}. 
On the client side, we use web browsers as apps because we need to parse the website's entire root page, and recursively fetch embedded objects. We also disable SPDY and caching.
Specifically, we use Android WebView\footnote{\url{http://developer.android.com/reference/android/webkit/WebView.html}}
for smartphone and smartglasses, and Android Wear Internet Browser\footnote{\url{https://play.google.com/store/apps/details?id=com.appfour.wearbrowser}}
for smartwatch. We cannot use WebView on the smartwatch because it is not supported.

\subsection{Evaluation Metrics}
We consider the amount of  \textit{downloaded data}, the \textit{data transfer time} and the \textit{energy consumption} as the main metrics for our analysis. 
To measure the \textit{downloaded data} and \textit{data transfer time}, we first run \texttt{tcpdump}  on each device to capture the packets for each experiment. We then extract the total number of bytes downloaded, as well as the time taken by  each phase of the HTTP/HTTPS connections by analyzing the TCP packet transmissions between the server and the client. 

We consider both the raw energy usage and the energy consumption normalized by the battery capacity of each device. Raw energy consumption is measured for every experiment with a Monsoon power monitor\footnote{\url{https://www.msoon.com/LabEquipment/PowerMonitor/}} directly connected to each device via USB. Energy is obtained by integrating the instantaneous power values calculated using current and voltage measurements
sampled at 0.2 ms time intervals. We carry out all the experiments with the battery fully charged to prevent additional charge current being drawn from the power monitor. Hence, the current drawn from the power monitor only performs the task associated with the experiment and background processes. The exact energy usage of the experiment is extracted by deducting the energy utilization of the background processes, and is approximately constant when the device is in the idle state.

\begin{figure}[t]
\centering  \vspace{-2mm}
\subfigure[LG G Watch R power profile for HTTP.]{\label{fig:timeHTTP}\includegraphics[width=0.5\textwidth]{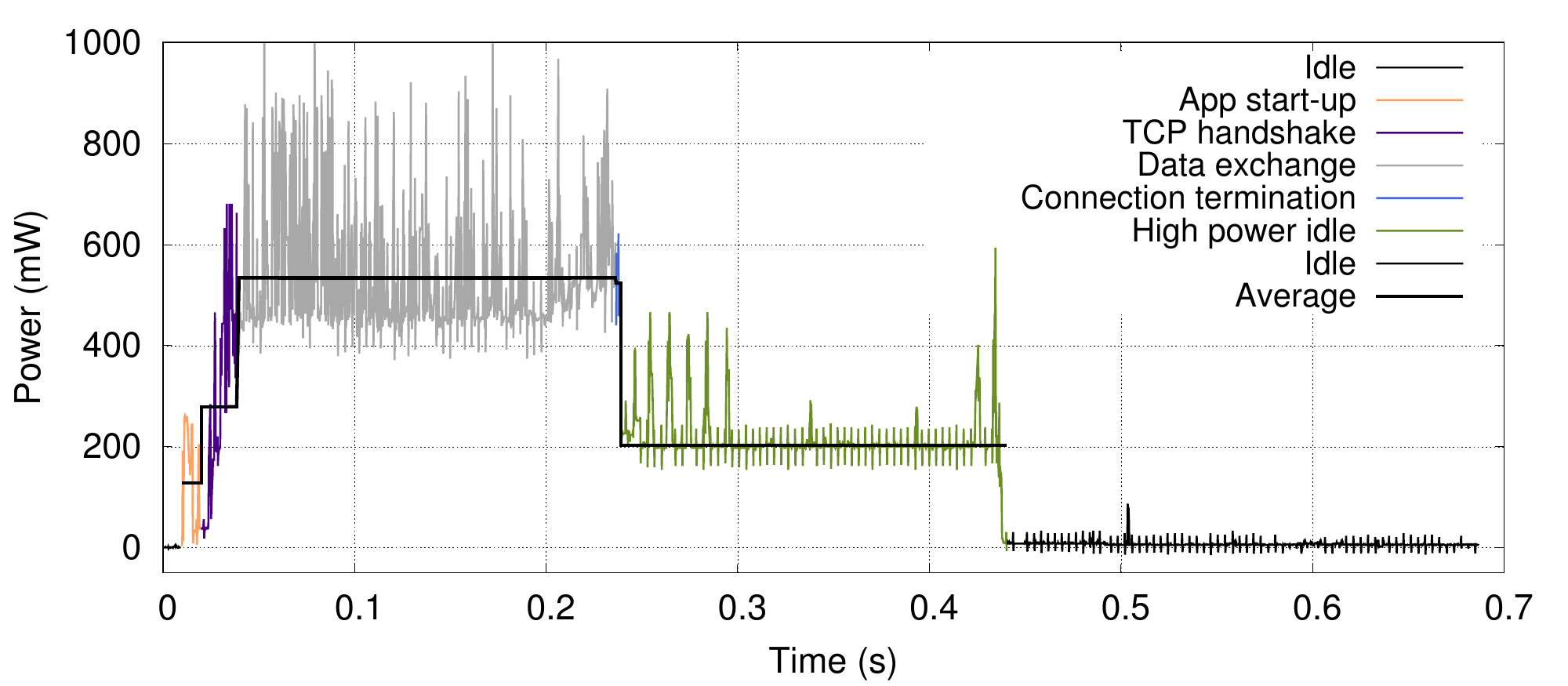}}\vspace{-2mm}
\subfigure[LG G Watch R power profile for HTTPS.]{\label{fig:timeHTTPS}\includegraphics[width=0.5\textwidth]{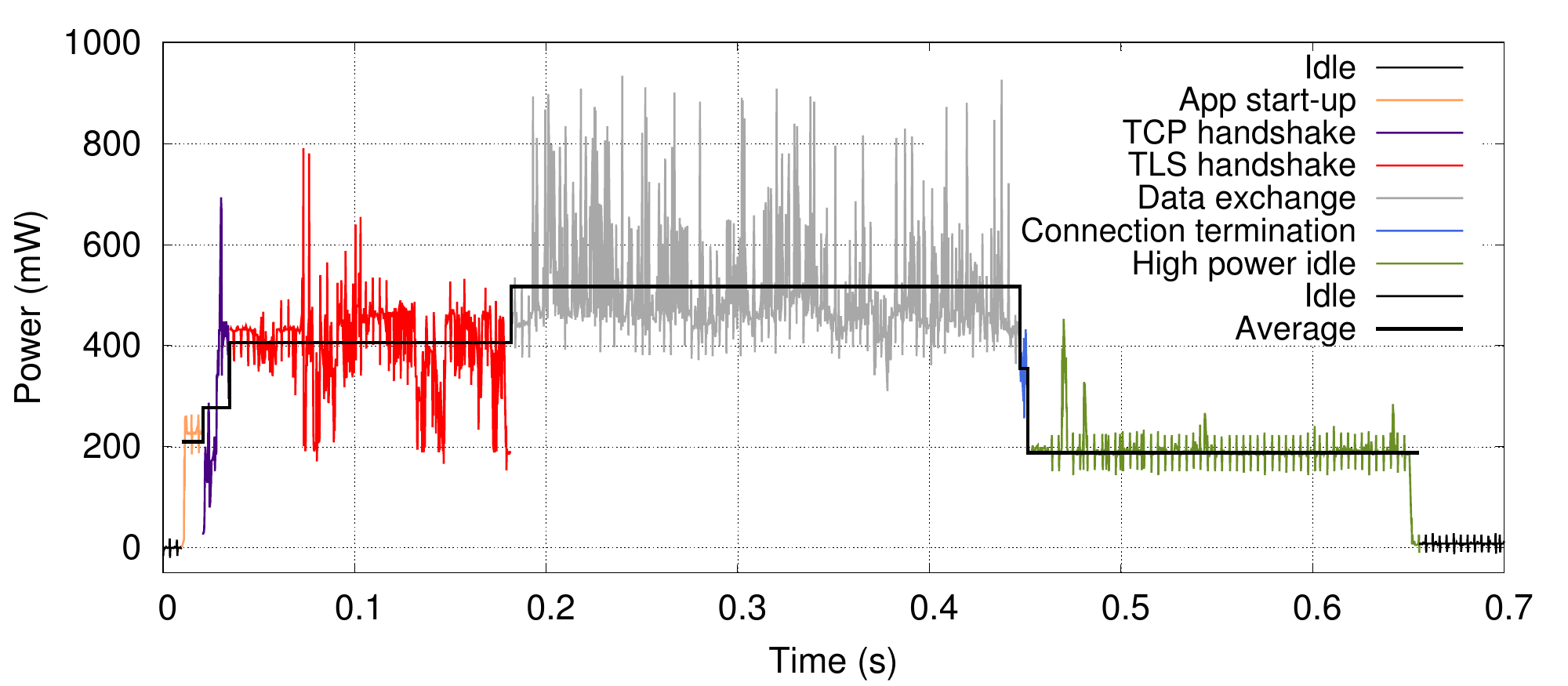}}\vspace{-2mm}
\caption{Energy consumption profile for the smartwatch LG G watch R when downloading a 500 KB file.}
\label{fig:downloadtime} 
 \vspace{-5mm}
\end{figure}

\begin{figure*}[tb]
\centering

\subfigure[Data transfer time.]{\label{fig:timAll}\includegraphics[width=0.246\textwidth]{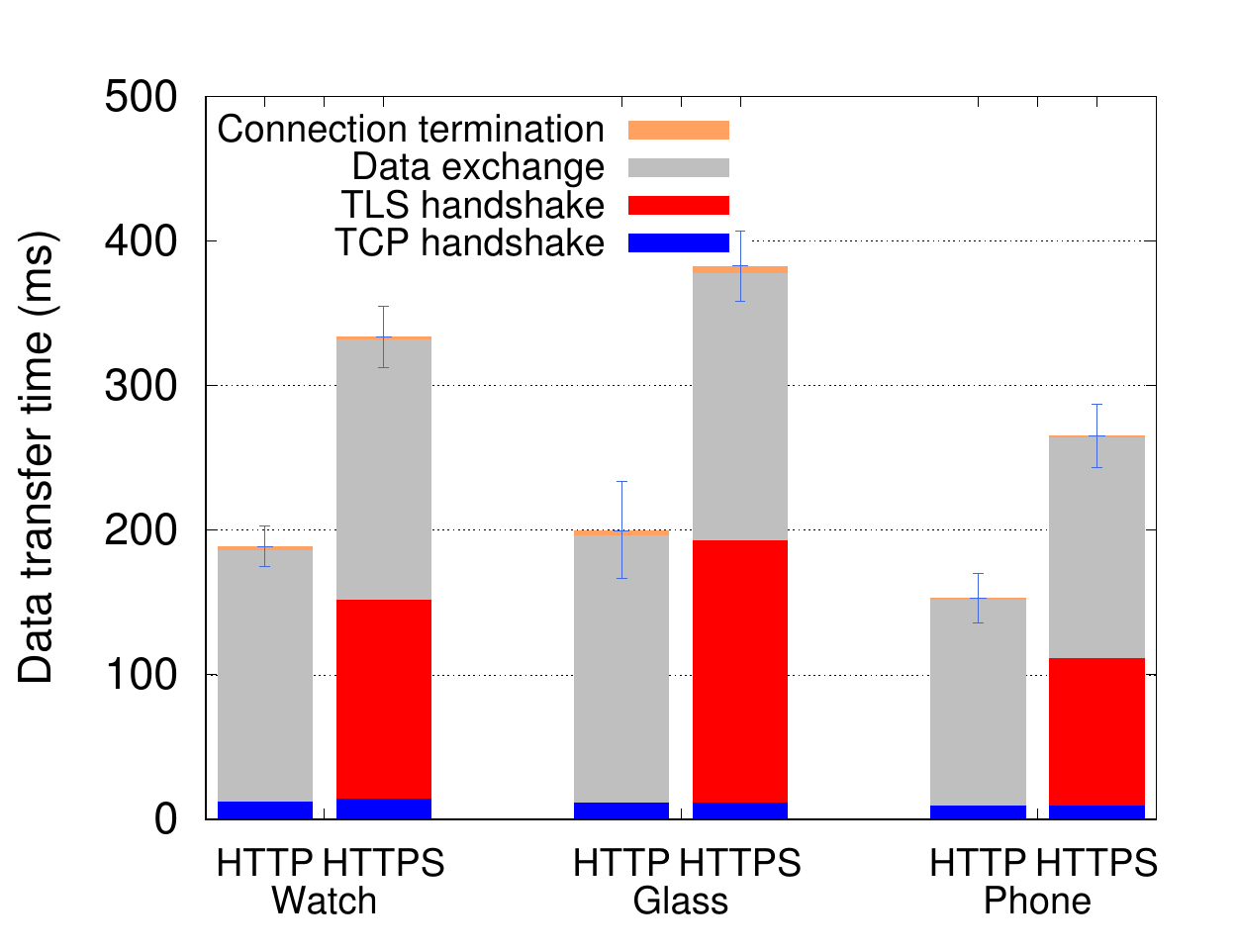}}
\subfigure[Energy consumption.]{\label{fig:energyAll}\includegraphics[width=0.246\textwidth]{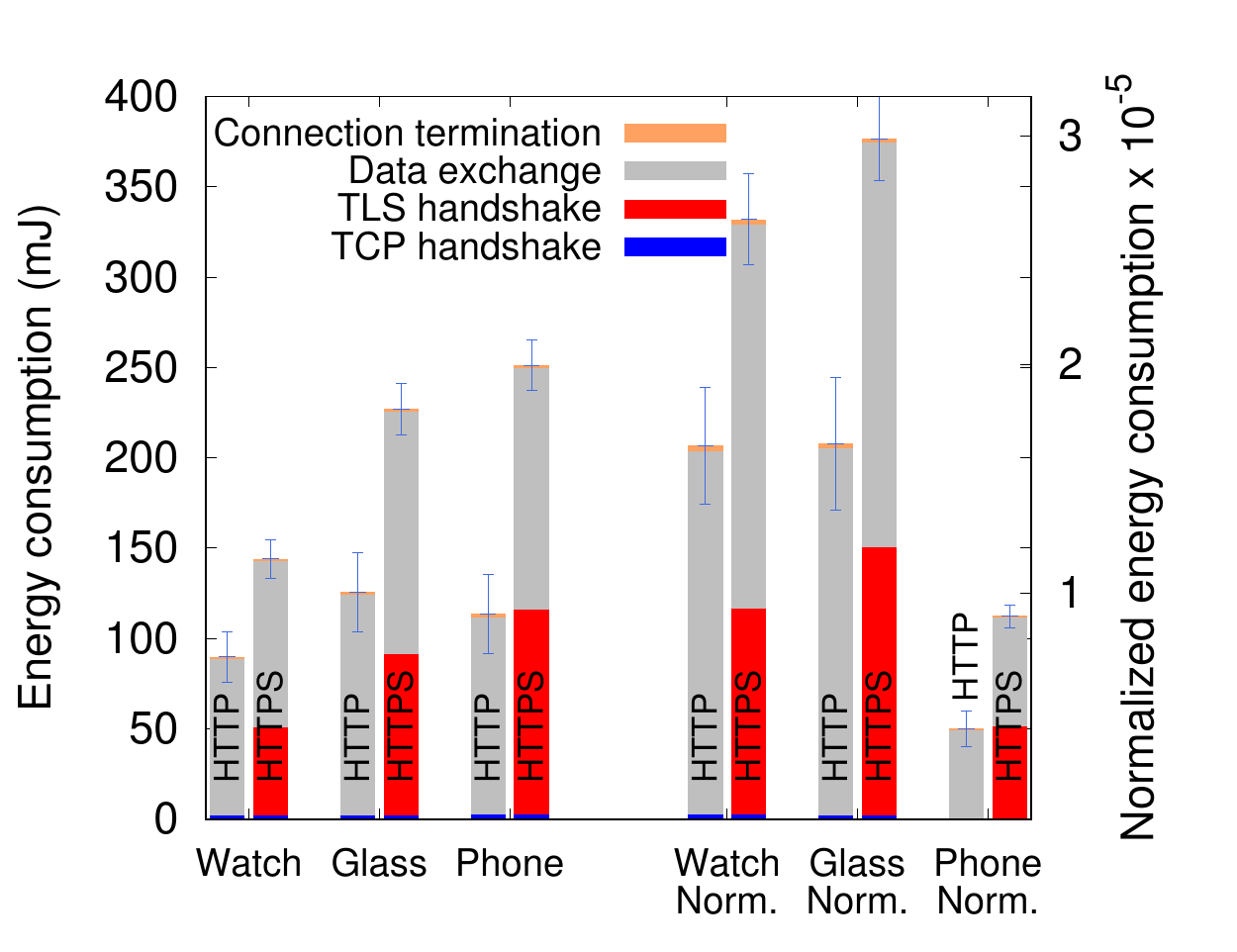}}
\subfigure[TLS phases duration.]{\label{fig:tlsphases}\includegraphics[width=0.246\textwidth]{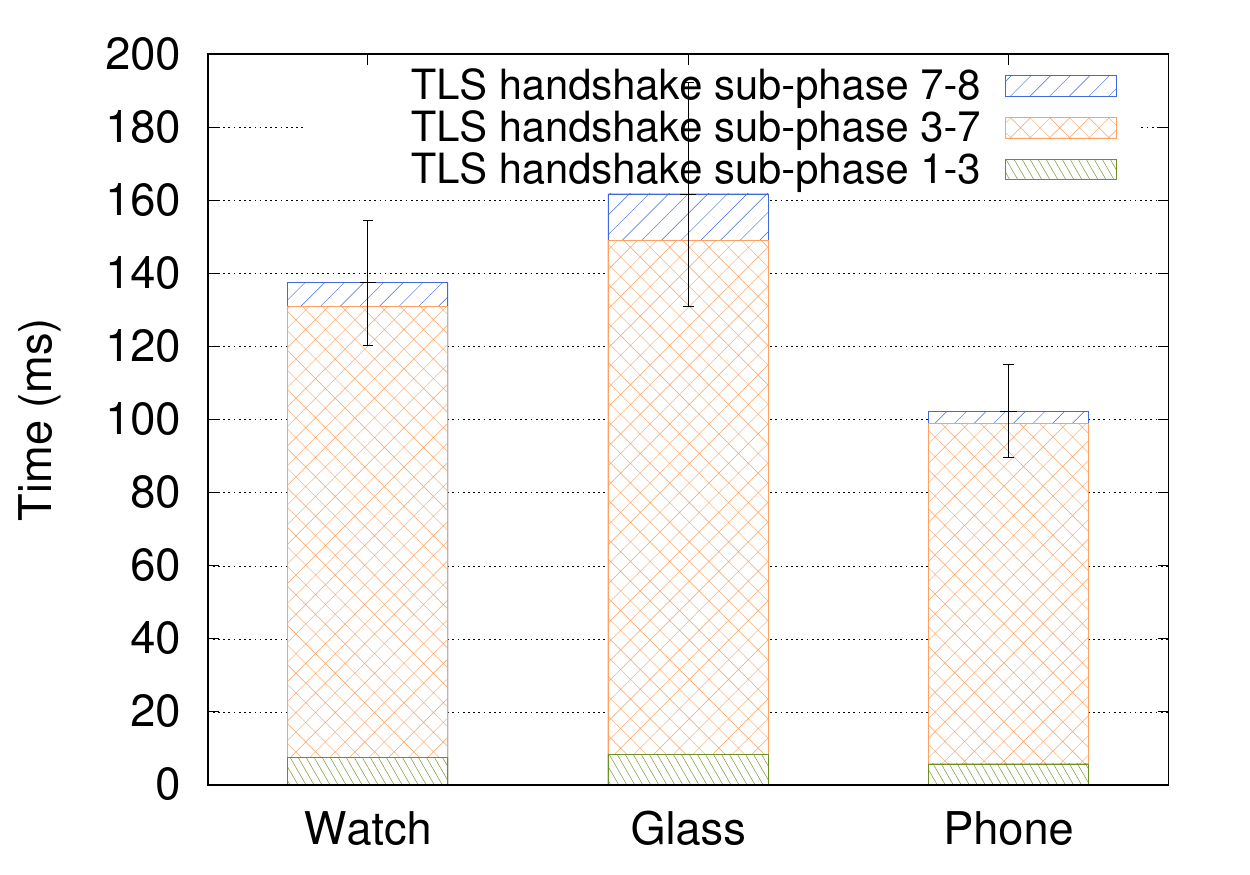}}
\subfigure[Wireless technology cost.]{\label{fig:energyBT}\includegraphics[width=0.246\textwidth]{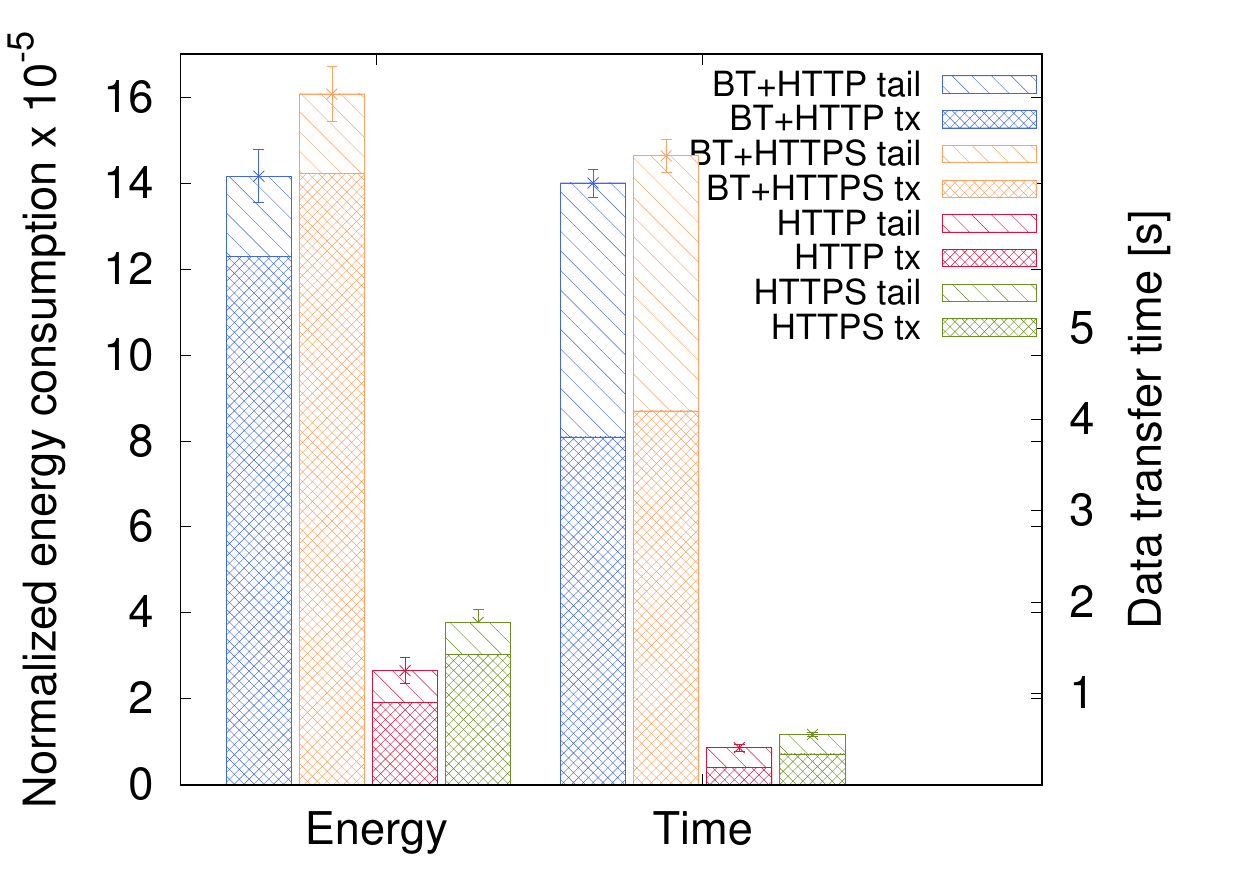}}

\caption{Cost of HTTP and HTTPS for a 500 KB file download (i.e., GET).}
\label{fig:filephase} 
\end{figure*}

\section{Controlled Testbed}
\label{sec:oneobject}
In this section, we present a set of experiments performed in a controlled testbed to characterize the impact of TLS on wearable devices. 
We first present a detailed analysis of a single file exchange for different sizes (\textsf{Exp1}), and then investigate the influence of TLS algorithms (\textsf{Exp2}).

\subsection{Understanding the Cost of TLS}

As an illustrative example, Figure~\ref{fig:downloadtime} shows the power consumption profile for the smartwatch when downloading a 500 KB file from the local web server. The profile shows the different phases of the data transfer, namely \textit{TCP handshake}, \textit{TLS handshake}, \textit{Data exchange} and \textit{Connection termination}. Additionally, we can observe \textit{Application start-up} and \textit{High power idle} phases in the power consumption profile.

We clearly observe that each phase has a different characteristic value for each metric. 
The highest power level is reached during the \textit{Data exchange} phase for both HTTP and HTTPS, on average 534.12 mW and 516.38 mW respectively. This cost is primarily due to the active radio transmitters/receivers. The power profile for the data exchange phase is closely correlated with the data transfer rate. By observing the elapsed time and bytes downloaded, we can infer a lower downlink data rate for HTTPS than HTTP  due to the processing overhead of encrypting data (i.e., 2.1 Mbps
 vs. 2.6 Mbps).
 Despite the lower average power, the total power consumption for data exchange is larger for HTTPS, because of the encryption and the longer data exchange time due to the combination of additional encryption bits and lower data rate.

TCP connection management (i.e, \textit{TCP handshake} and \textit{Connection termination}) consumes very little energy as it is composed of only a few packets. Energy consumption for \textit{Application start-up} is not negligible, although there is no data exchanged during this period. \textit{High power idle} profile ($\sim$200 mW for $\sim$200 ms) is similar for both HTTP and HTTPS as it is independent from the higher layer protocols.  

As explained in the background section, \textit{TLS handshake} is the additional phase in HTTPS where the client and server agree upon the TLS attributes for the session and exchange keys for payload encryption/decryption. Although the TLS handshake generates limited data, it has a significant energy footprint as 
this phase takes about 147 ms and 405.22 mW on average; we further examine the performance details in the following paragraphs.

Figure~\ref{fig:timAll} and \ref{fig:energyAll} present the average values of \textit{data transfer time} and \textit{energy consumption} for each phase of the HTTP/HTTPS connection respectively, while Table~\ref{table:trace} reports the \textit{downloaded data} in bytes for each phase. Overall as expected, we observe that TLS increases the \textit{data transfer time} by  $\sim$75\% to $\sim$90\%, the amount of \textit{downloaded data} by $\sim$2\%  and the \textit{energy consumption} by $\sim$70\% to $\sim$100\%  for all devices when downloading 500KB file. 

\begin{table}[tb]
\scriptsize
\centering 
\begin{tabu}{ccccc}
\tabucline[0.8pt]{-}
\noalign{\smallskip}
\textbf{Protocol}&\textbf{TCP}&\textbf{TLS}&\textbf{Data}&\textbf{Connection}\\
&\textbf{hand.}&\textbf{hand.}&\textbf{down.}&\textbf{term.}\\\hline\hline\noalign{\smallskip}
HTTP&74&-&523212&66\\
HTTPS&74&5421&524390&132\\
\noalign{\smallskip}
\tabucline[0.8pt]{-}
\end{tabu}
\caption{Downloaded data amount (Bytes) for different categories.}
\label{table:trace}
\vspace{-4 mm}
\end{table}

Also as expected, the large file size used in this experiment makes the impact of other phases negligible on the total amount of data exchanged as can be seen from Table~\ref{table:trace}. 
The smartphone provides the best performance in terms of \textit{data transfer time} followed by smartwatch and smartglass (cf. Figure ~\ref{fig:timAll}). This reflects the hardware capabilities of the different devices, which in turn influences the \textit{energy consumption} as shown in Figure~\ref{fig:energyAll}. On the one hand, better hardware components consume more power as we observe on the left-hand side (from the middle) of Figure~\ref{fig:energyAll}. On the other hand, as the battery capacity of smartphone is larger than wearable counterparts, the relative energy consumption is more significant on wearables as shown on the right-hand side (from the middle) of Figure~\ref{fig:energyAll}.

\begin{figure*}[t]
\centering \vspace{-4mm}
\subfigure[Data transfer time.]{\label{fig:timeHTTPS/HTTP}\includegraphics[width=0.245\textwidth]{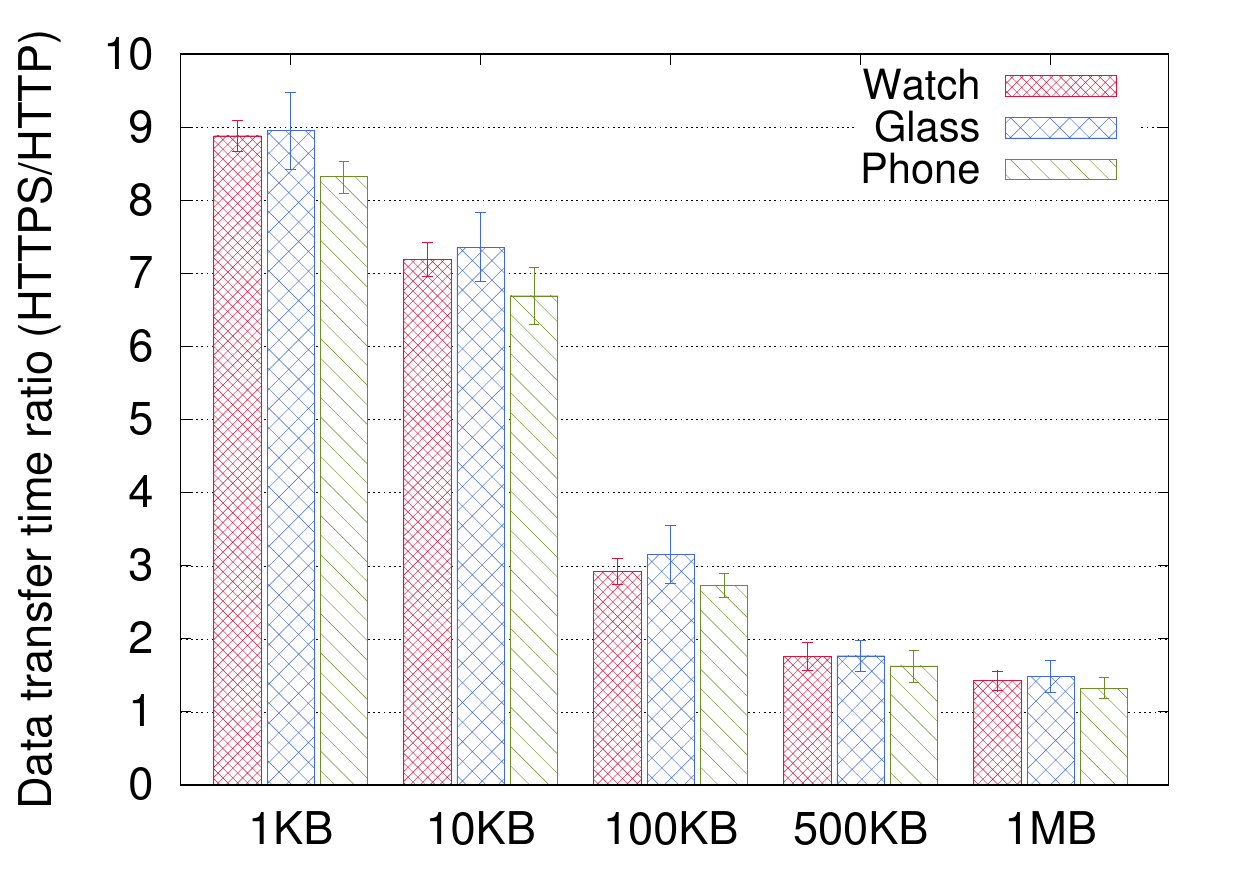}}
\subfigure[Data usage.]{\label{fig:bytesHTTPS/HTTP}\includegraphics[width=0.245\textwidth]{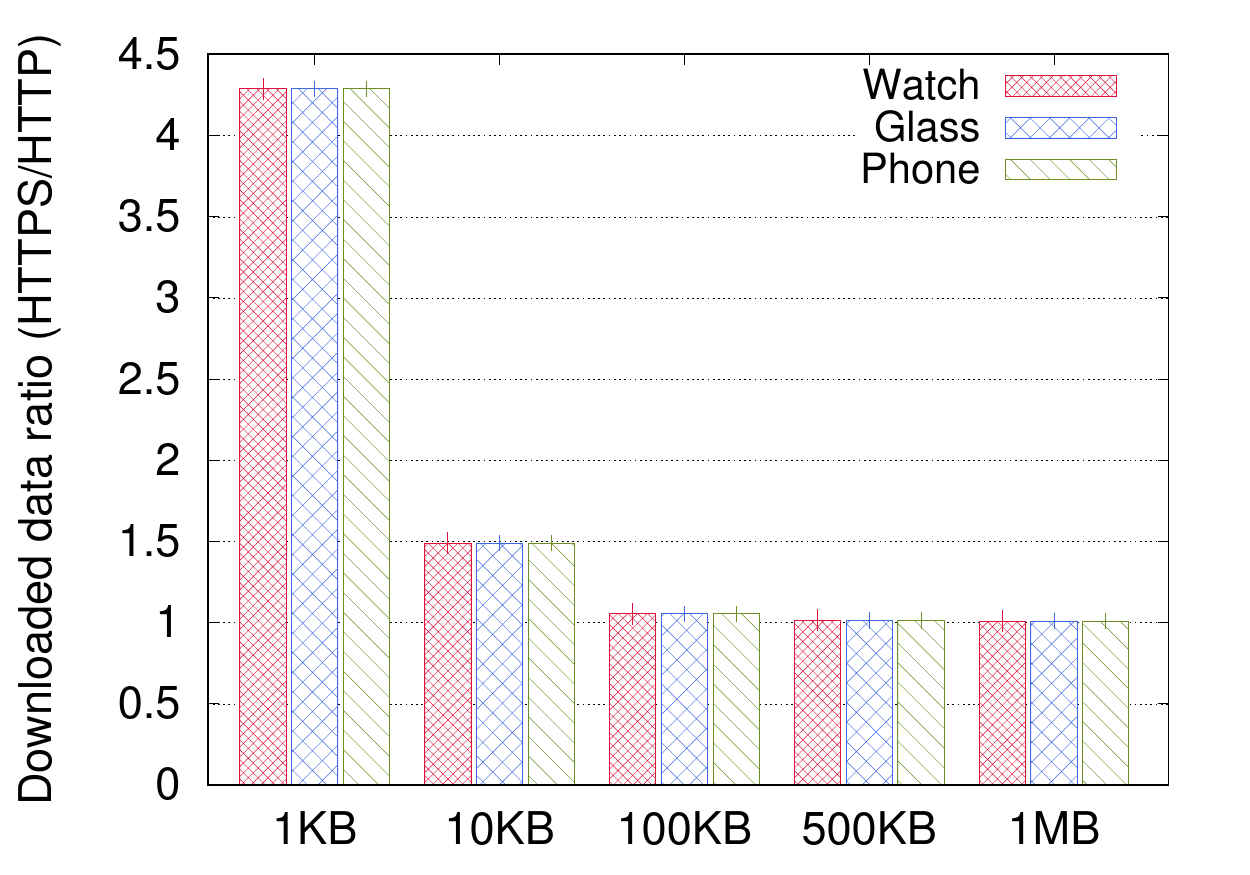}}
\subfigure[Energy consumption.]{\label{fig:energyHTTPS/HTTP}\includegraphics[width=0.245\textwidth]{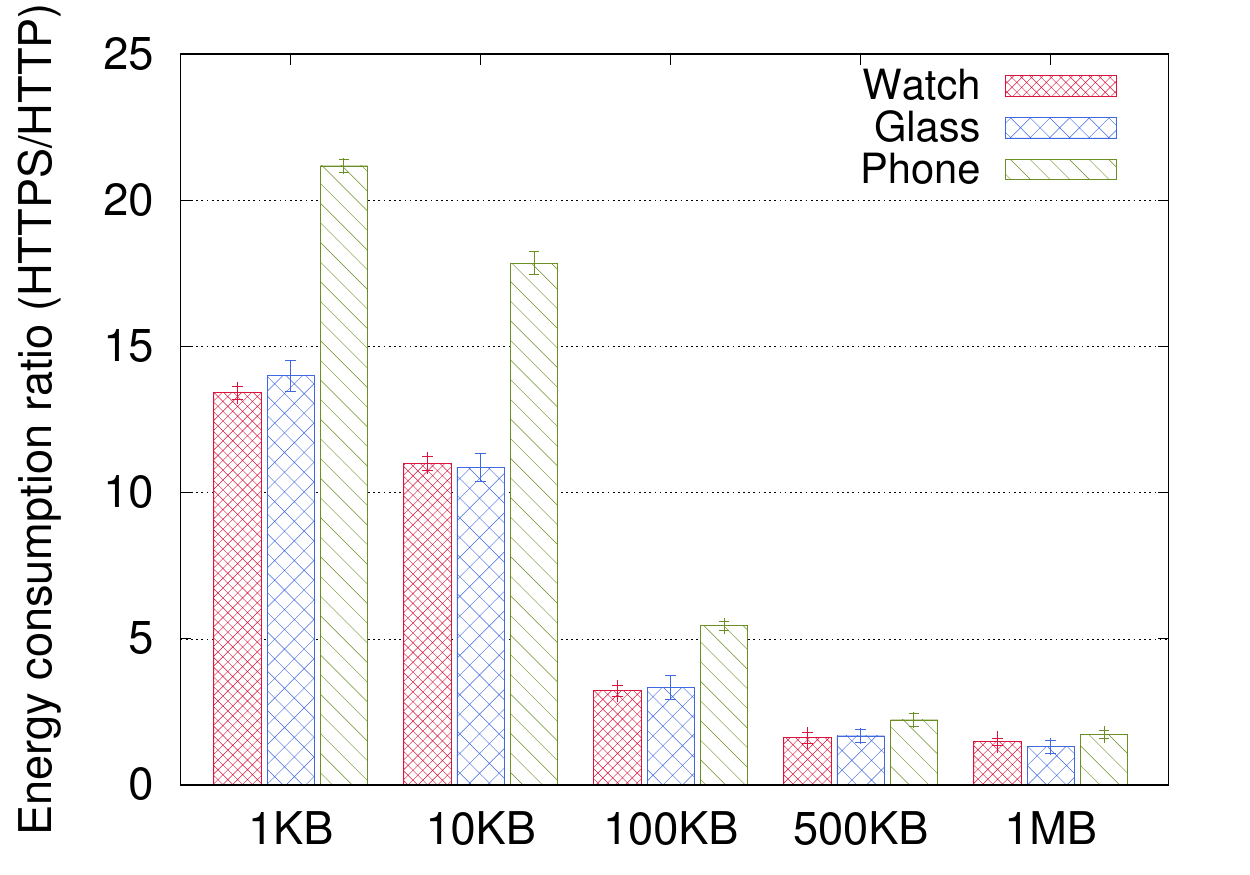}}
\subfigure[Normalized energy.]{\label{fig:normalized}\includegraphics[width=0.245\textwidth]{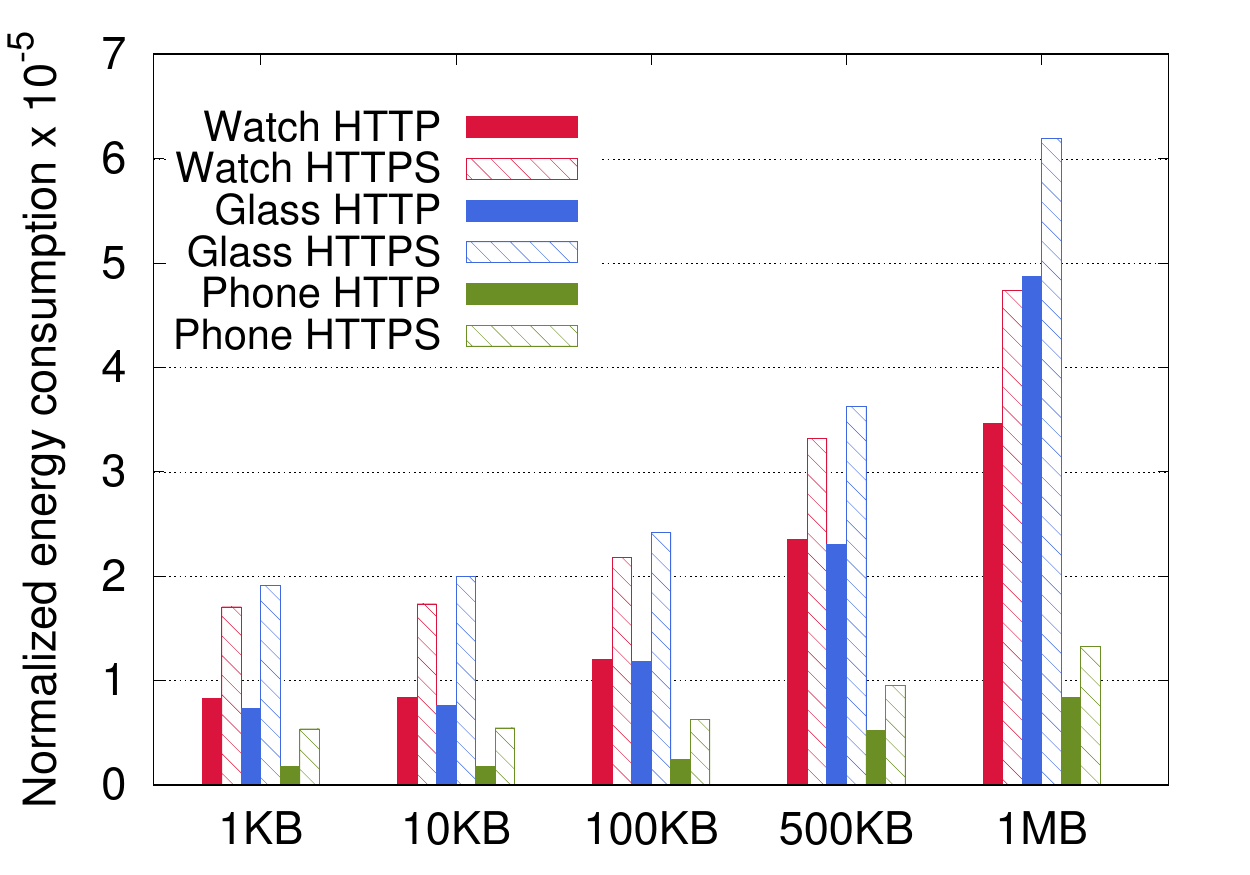}}
\caption{Impact of file size on data transfer time, data downloaded and energy consumption.}
\label{fig:filesize} 
\vspace{-5mm}
\end{figure*}

We further analyze the TLS handshake process in detail for all devices in Figure~\ref{fig:tlsphases}. We make two main observations. First, the different duration of the TLS handshake across devices reflects their hardware characteristics.
Second, certificate validation and key computation on the client side (\textit{TLS handshake} sub-phase 3-7 explained in the background section)
is taking $\sim$87\% of the total time of \textit{TLS handshake}.

\begin{figure*}[t]
\centering 

\subfigure[Data transfer time.]{\label{fig:ssltimeHTTPS/HTTP}\includegraphics[width=0.245\textwidth]{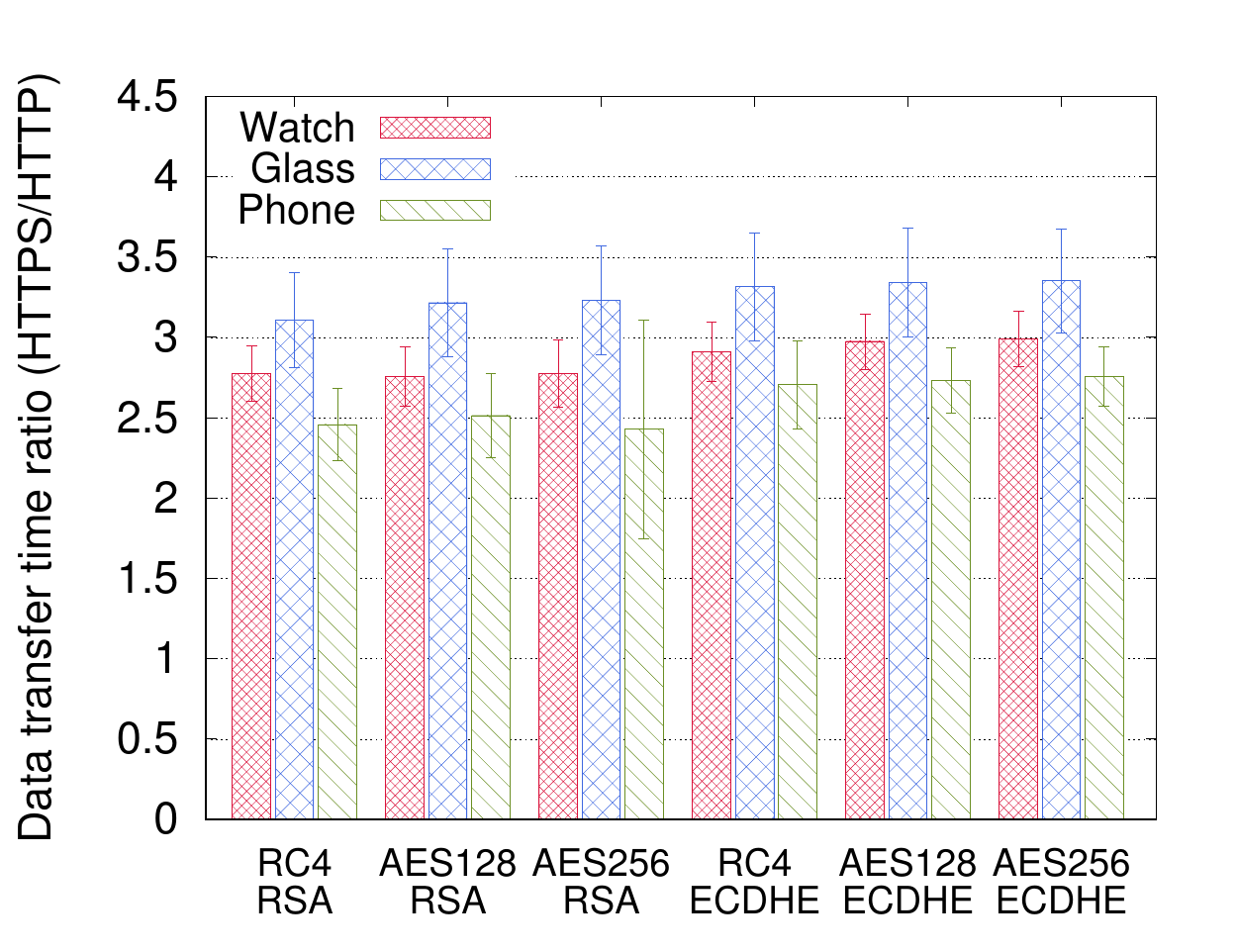}}\vspace{-2mm}
\subfigure[Energy consumption.]{\label{fig:sslenergyHTTPS/HTTP}\includegraphics[width=0.245\textwidth]{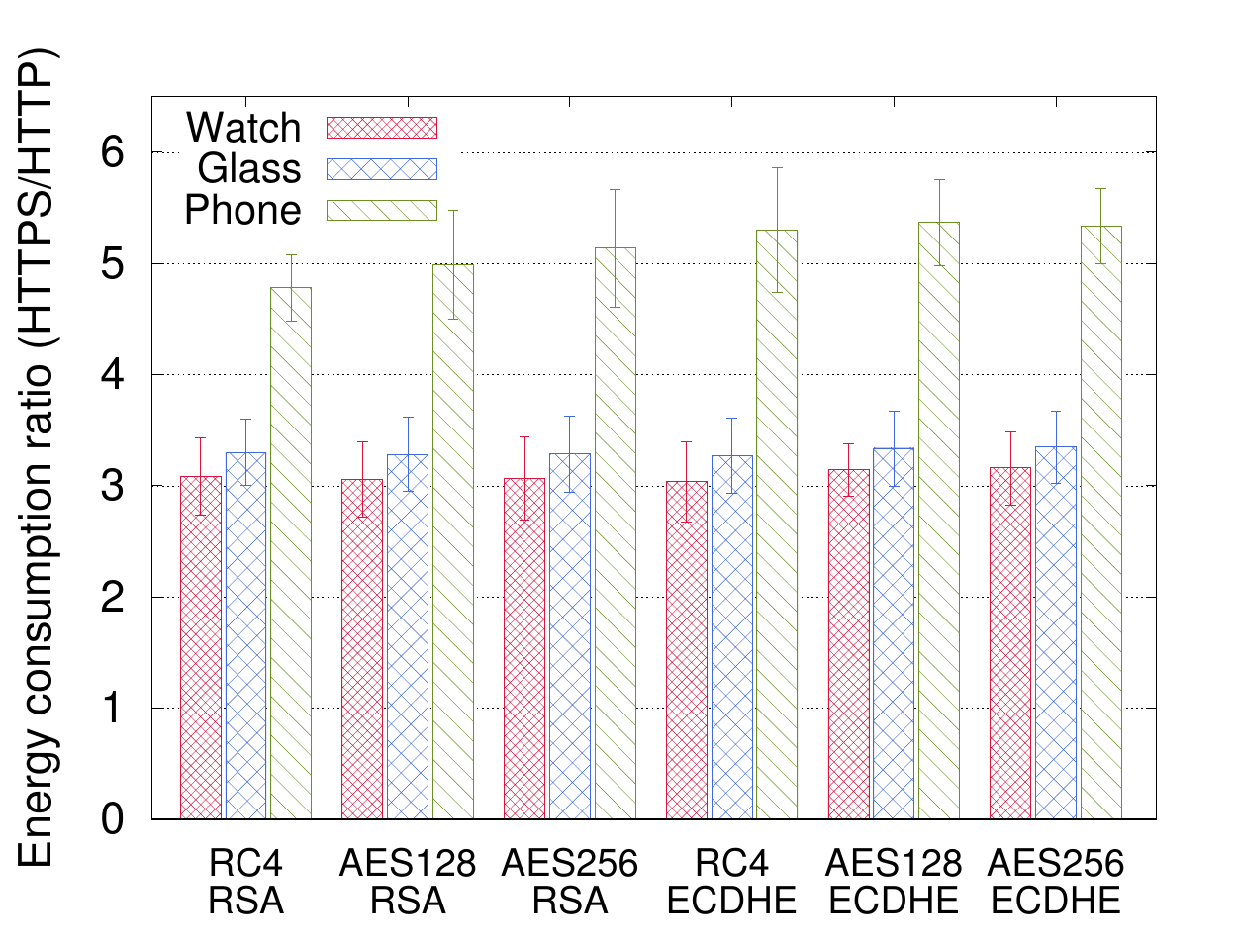}}
\subfigure[Data transfer time.]{\label{fig:ssltimeraw}\includegraphics[width=0.245\textwidth]{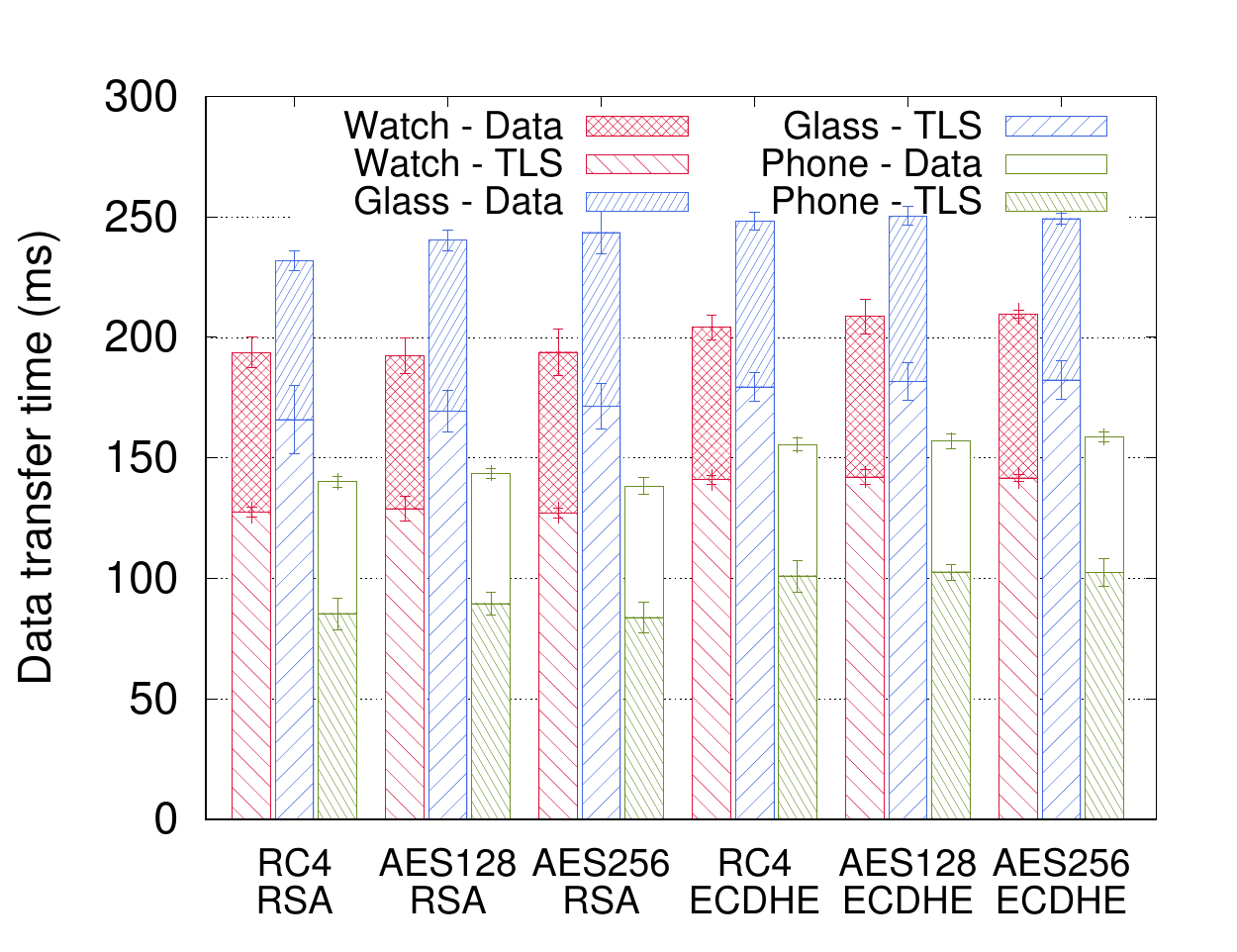}}
\subfigure[Energy consumption.]{\label{fig:sslenergyraw}\includegraphics[width=0.245\textwidth]{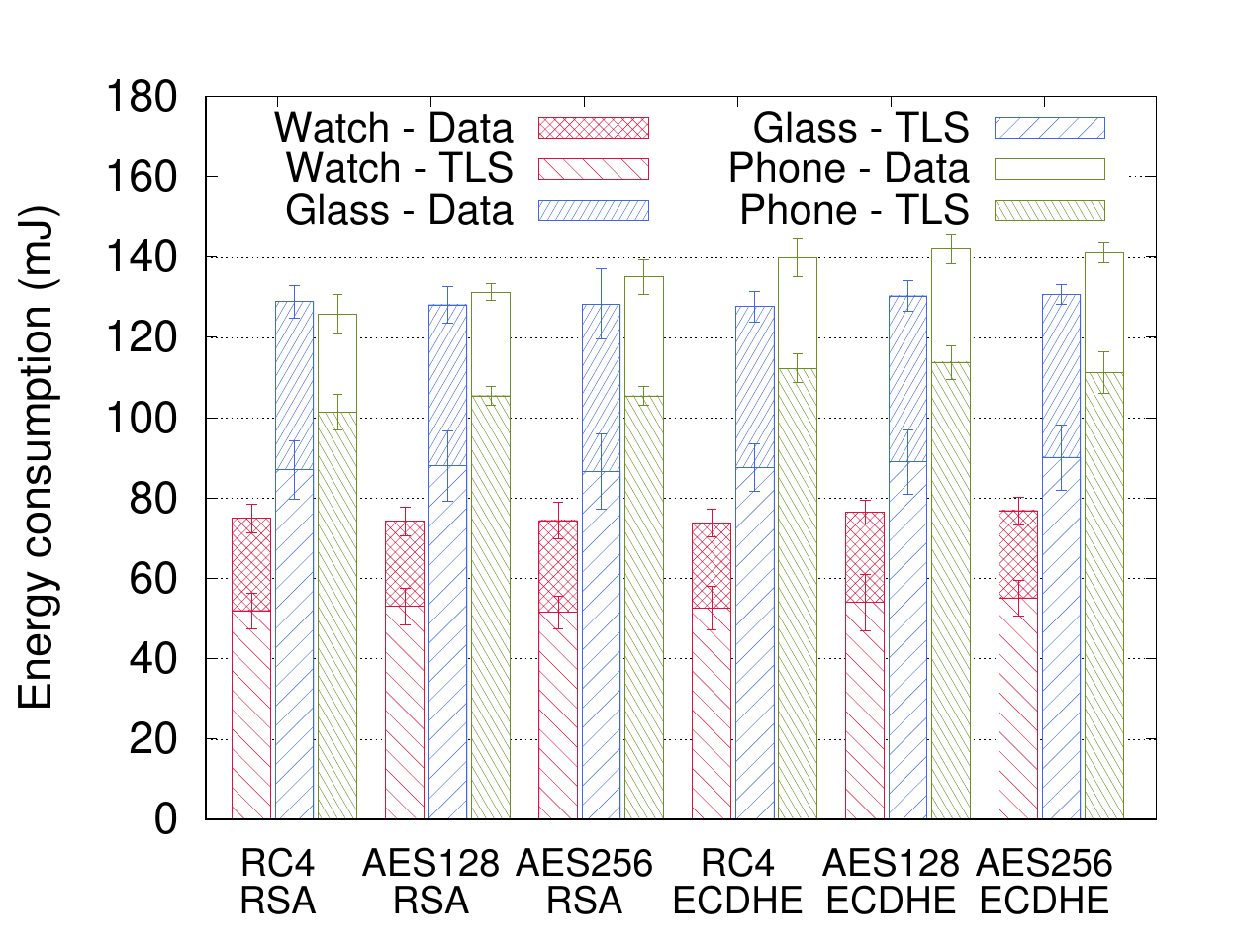}}
\caption{Impact of TLS parameters on data transfer time and energy consumption.}
\label{fig:sslratioparameters} 
\vspace{-5mm}

\end{figure*}

\begin{figure*}[tb]
\centering  \vspace{-4mm}
\subfigure[Watch]{\label{fig:bar}\includegraphics[width=0.245\textwidth]{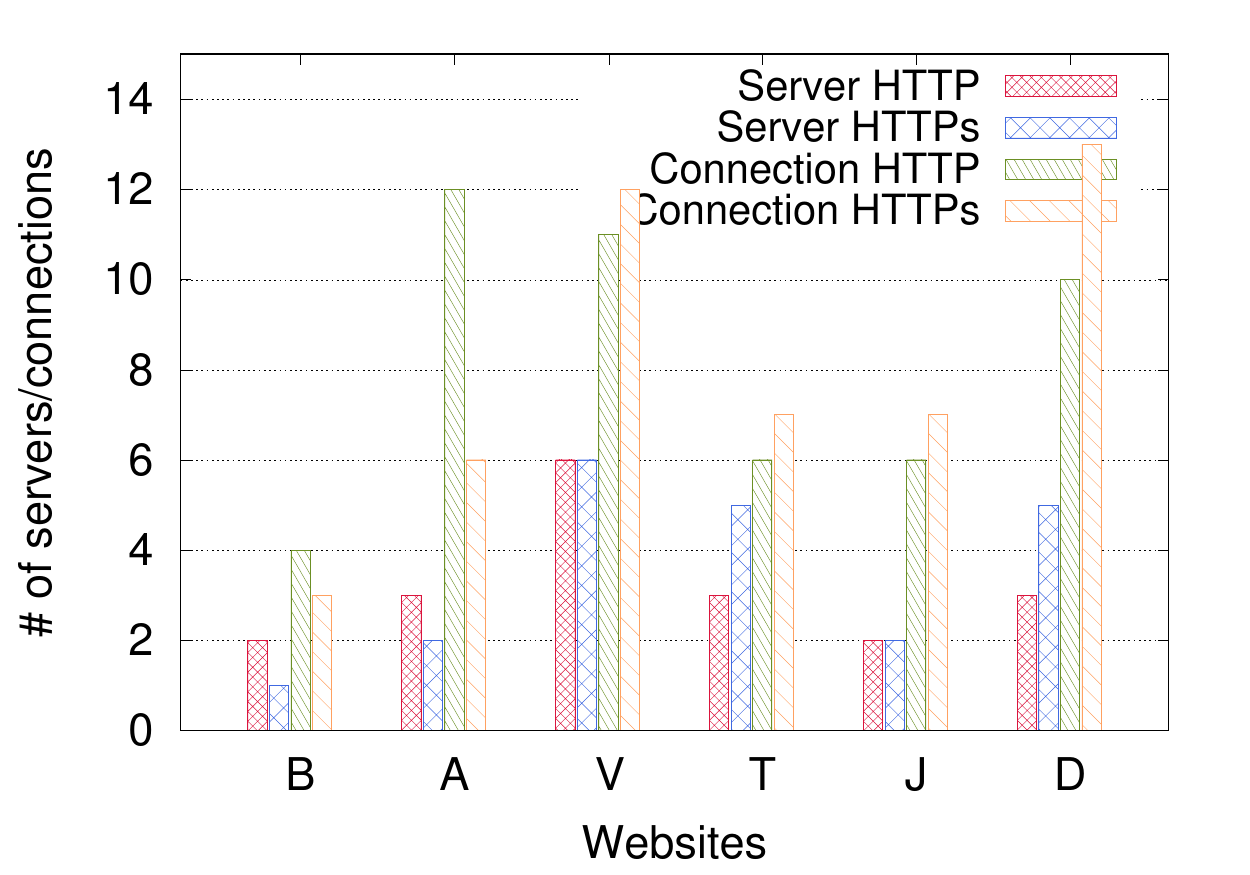}}
\subfigure[Glass]{\label{fig:energybarc}\includegraphics[width=0.245\textwidth]{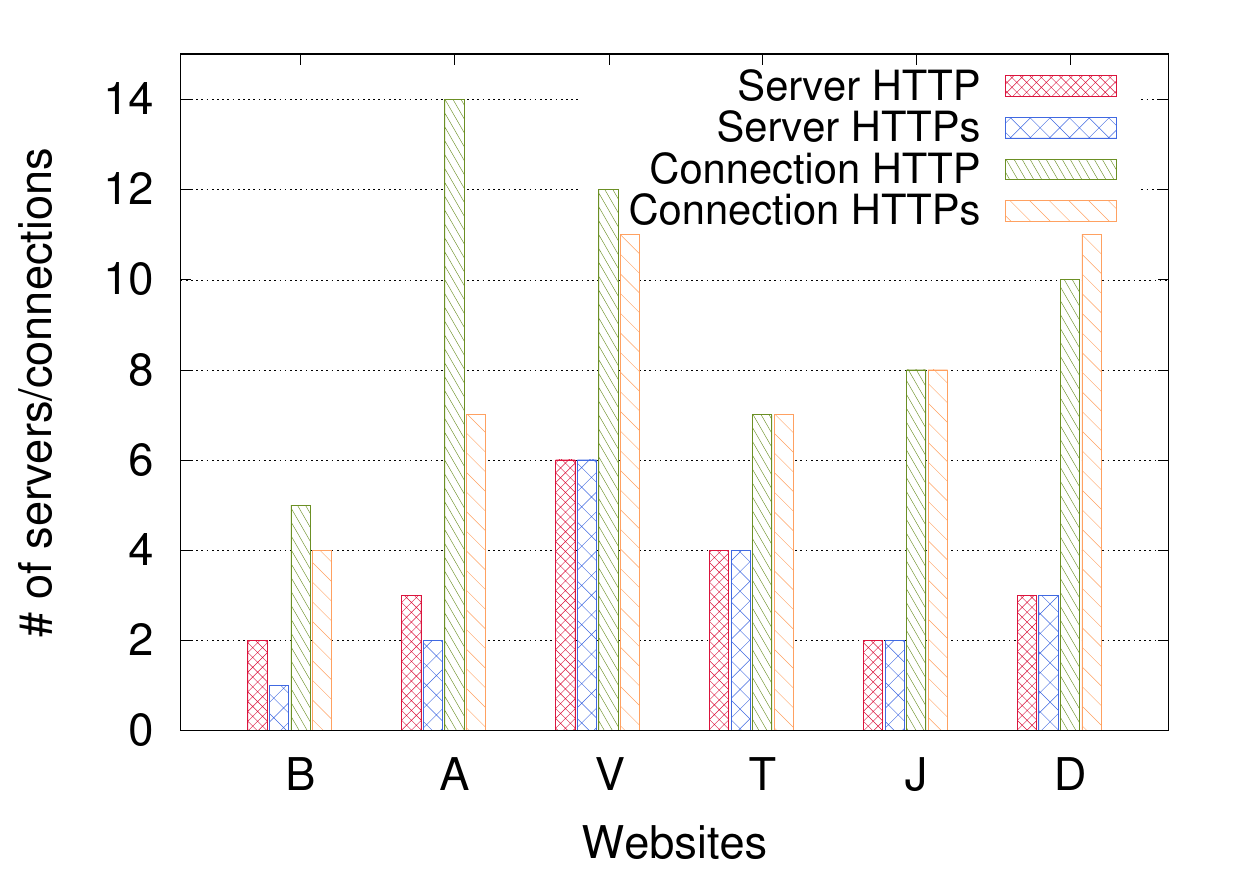}}
\subfigure[Phone]{\label{fig:barb}\includegraphics[width=0.245\textwidth]{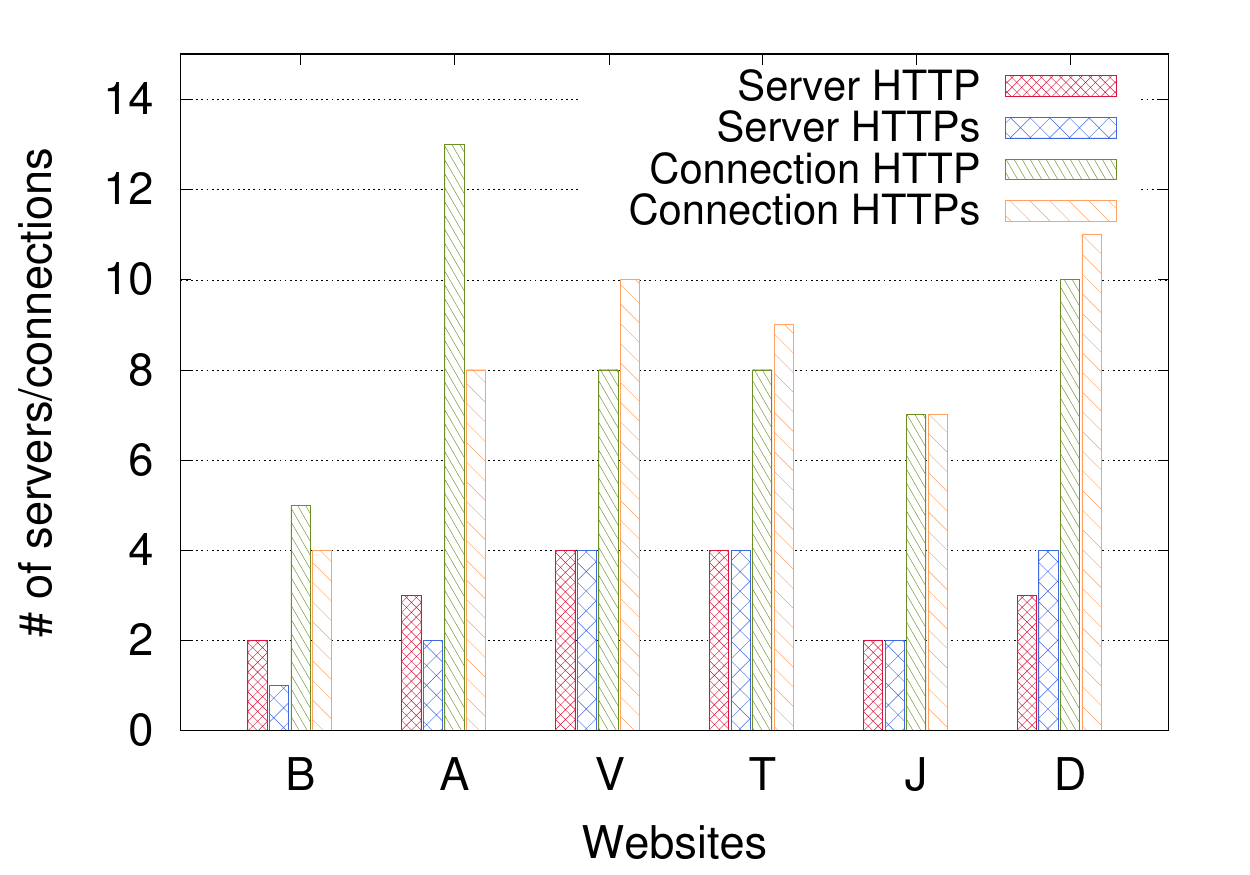}} 
\subfigure[Device RTT]{\label{fig:rtt}\includegraphics[width=0.245\textwidth]{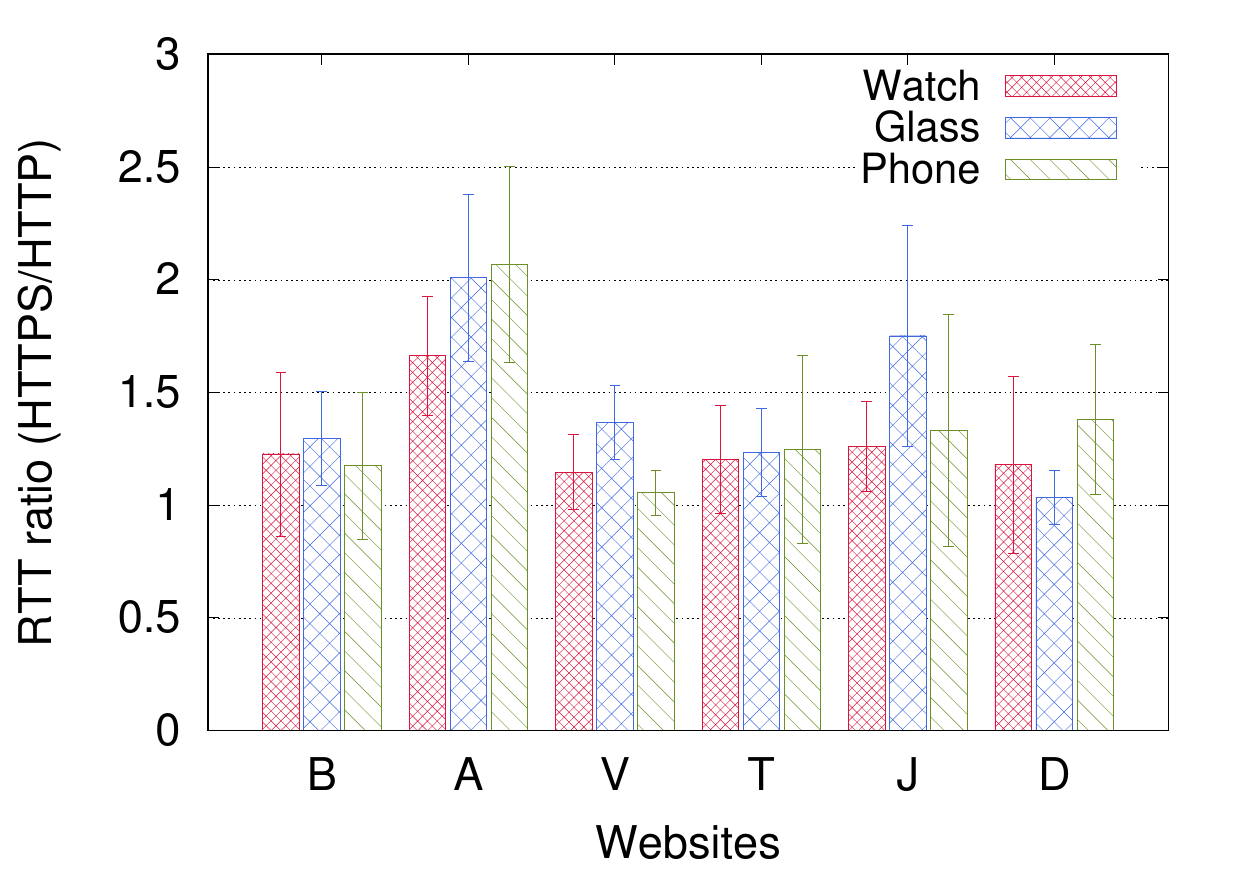}} 
\caption{(a)-(b)-(c) Analysis of the number of connections/servers involved in data exchange between mobile devices and Internet websites. Refer to Table~\ref{tab:websites} for website identifiers. (d) Impact of RTT on $\frac{\text{HTTPS}}{\text{HTTP}}$ ratio. }  \vspace{-5mm}
\label{fig:internetbar} 
\end{figure*}

\subsection{Impact of the Wireless Communication Technology}

Bluetooth communication is utilized if smartwatch apps leverage the smartphone to connect to the Internet, as this is the chosen method for most current applications. Figure ~\ref{fig:energyBT} shows the smartwatch \textit{energy consumption} which is normalized by 100\% battery capacity and \textit{data transfer time} for Bluetooth, and compare them to HTTP and HTTPS communication over WiFi.
The \textit{energy consumption} of this additional data exchange between smartwatch and smartphone is $\sim$8 times higher than HTTP and $\sim$5 times higher than HTTPS via WiFi in smartwatch. And also, Bluetooth relaying adds $\sim$40\% more \textit{data transfer time} than HTTPS via WiFi in smartwatch. WiFi guarantees the minimal power/throughput ratio~\cite{friedmanpower}, and is thus more power efficient than Bluetooth for file transfer. Also, the Bluetooth transfer significantly increases the transmission delay because of the lower transmission rate compared to WiFi. 

\subsection{Impact of the Transaction File Size} 
The effect of TLS protocol selection on the varying sizes of downloaded files is shown in Figure~\ref{fig:filesize}. Overall, the $\frac{\text{HTTPS}}{\text{HTTP}}$ ratio is comparable across devices and decreases as the file size increases for all metrics: the larger the file the more the cost of TLS handshake is amortized. 

The ratio of \textit{data transfer time} ($\frac{\text{HTTPS}}{\text{HTTP}}$) is highest in smartglasses followed by smartwatch and smartphone for all file sizes (Figure \ref{fig:timeHTTPS/HTTP}). However, the magnitude of the \textit{download bytes} ratio ($\frac{\text{HTTPS}}{\text{HTTP}}$) is almost identical for all three devices as they use the same TLS parameters (Figure \ref{fig:bytesHTTPS/HTTP}). Finally, Figure~\ref{fig:energyHTTPS/HTTP} and \ref{fig:normalized} show the impact of TLS is higher on devices with better hardware components, but has a lower relative impact as those devices are also equipped with larger-capacity batteries. Also, despite the higher $\frac{\text{HTTPS}}{\text{HTTP}}$ \textit{energy consumption} ratio for small files, the relative impact of those protocols is higher for large files because of the longer transfer duration and encryption/decryption done for more bytes.

\subsection{Impact of Cryptographic Algorithms}
\label{sec:ssl}

The results of \textit{data transfer time} ratio and \textit{energy consumption} ratio obtained by running different combinations of cryptographic algorithms are shown in Figure \ref{fig:ssltimeHTTPS/HTTP} and \ref{fig:sslenergyHTTPS/HTTP}, while Table \ref{table:rawssl} shows the \textit{downloaded bytes} ratio. 
Moreover, Figure \ref{fig:ssltimeraw} and \ref{fig:sslenergyraw} show the effect of changing TLS parameters on \textit{Data exchange} and \textit{TLS handshake} phases.
 As expected, various cryptographic algorithms have different impacts on HTTPS performance for all devices. Overall, changing cryptographic parameters are less significant than other parameters (e.g., file size).

\begin{table}[tb]
\scriptsize
\centering 
\begin{tabu}{cccccc}
\tabucline[0.8pt]{-}
\noalign{\smallskip}
\textbf{RSA+}&\textbf{RSA+}&\textbf{RSA+}&\textbf{ECDH+}&\textbf{ECDH+} &\textbf{ECDH+}\\
\scriptsize{\textbf{RC4}}&\scriptsize{\textbf{AES128}}&\scriptsize{\textbf{AES256}}&\scriptsize{\textbf{RC4}}&\scriptsize{\textbf{AES128}}&\scriptsize{\textbf{AES256}}
\\\hline\hline\noalign{\smallskip}
1.0515&1.0524&1.0524&1.0548&1.0557&1.0557\\
\noalign{\smallskip}
\tabucline[0.8pt]{-}
\end{tabu}
\caption{$\frac{\text{HTTPS}}{\text{HTTP}}$ ratio of download data amount (Bytes) for different TLS parameters.}
\label{table:rawssl}
\vspace{-5.5mm}
\end{table}

Using ECDHE  instead  of RSA leads to an increase in the \textit{downloaded bytes}, \textit{data transfer time} and \textit{energy consumption}. 
As mentioned earlier, the key exchange algorithm has a significant impact on the key exchange phase for all the three devices. For example, changing key exchange algorithm from (RSA+RC4) to (ECDHE+RC4) leads to an $\sim$7\% increase in \textit{downloaded bytes}, $\sim$10\% increase in \textit{data transfer time} and $\sim$1\% increase in \textit{energy consumption} on the smartwatch's  key exchange phase. This is because ECDHE uses more complex mathematical operations than RSA to perform the encryption/decryption. 

The choice of bulk cipher algorithm in TLS does not have a significant impact on the metrics. For example, changing bulk cipher algorithm from (RSA + RC4) to (RSA + AES128) leads to$\sim$0.1\% increase in \textit{downloaded bytes}, $\sim$1\% increase in \textit{data transfer time} and $\sim$1\% increase in \textit{energy consumption} on the smartwatch's application data transfer phase.

\section{Internet experiments}
\label{sec:internet}

\begin{figure*}[ht]
\centering 
\subfigure[]{\label{fig:scatsize}\includegraphics[width=0.29\textwidth]{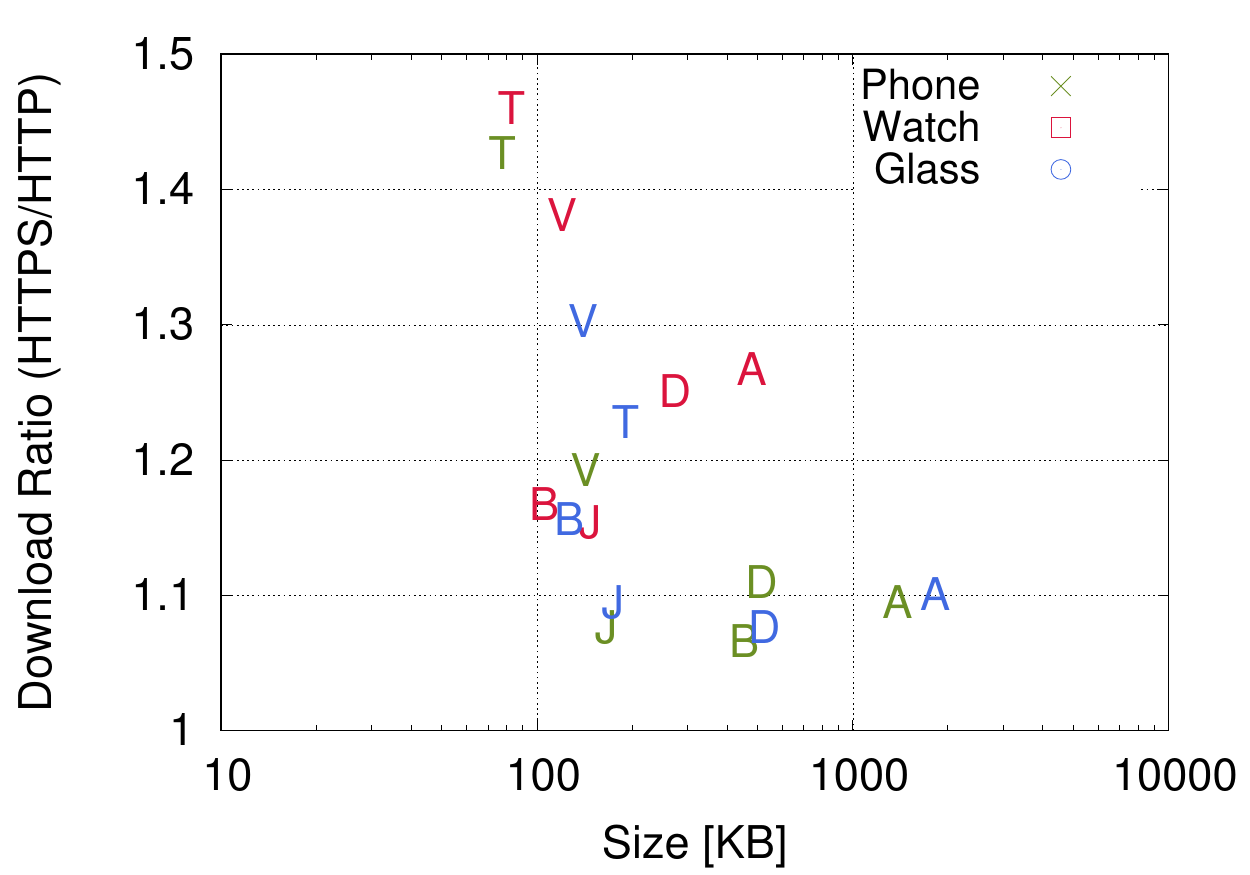}}
\subfigure[]{\label{fig:scattime}\includegraphics[width=0.29\textwidth]{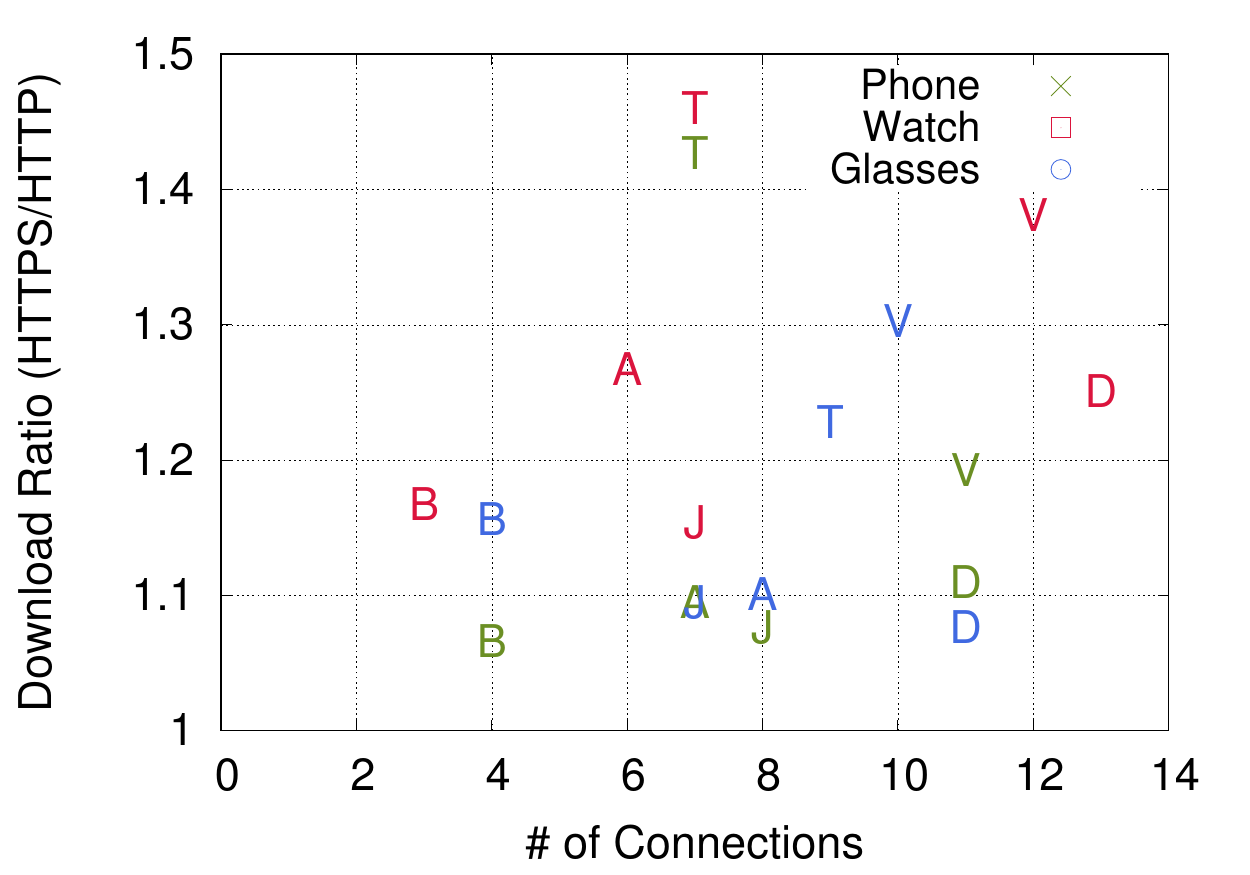}}
\subfigure[]{\label{fig:scatenergy}\includegraphics[width=0.29\textwidth]{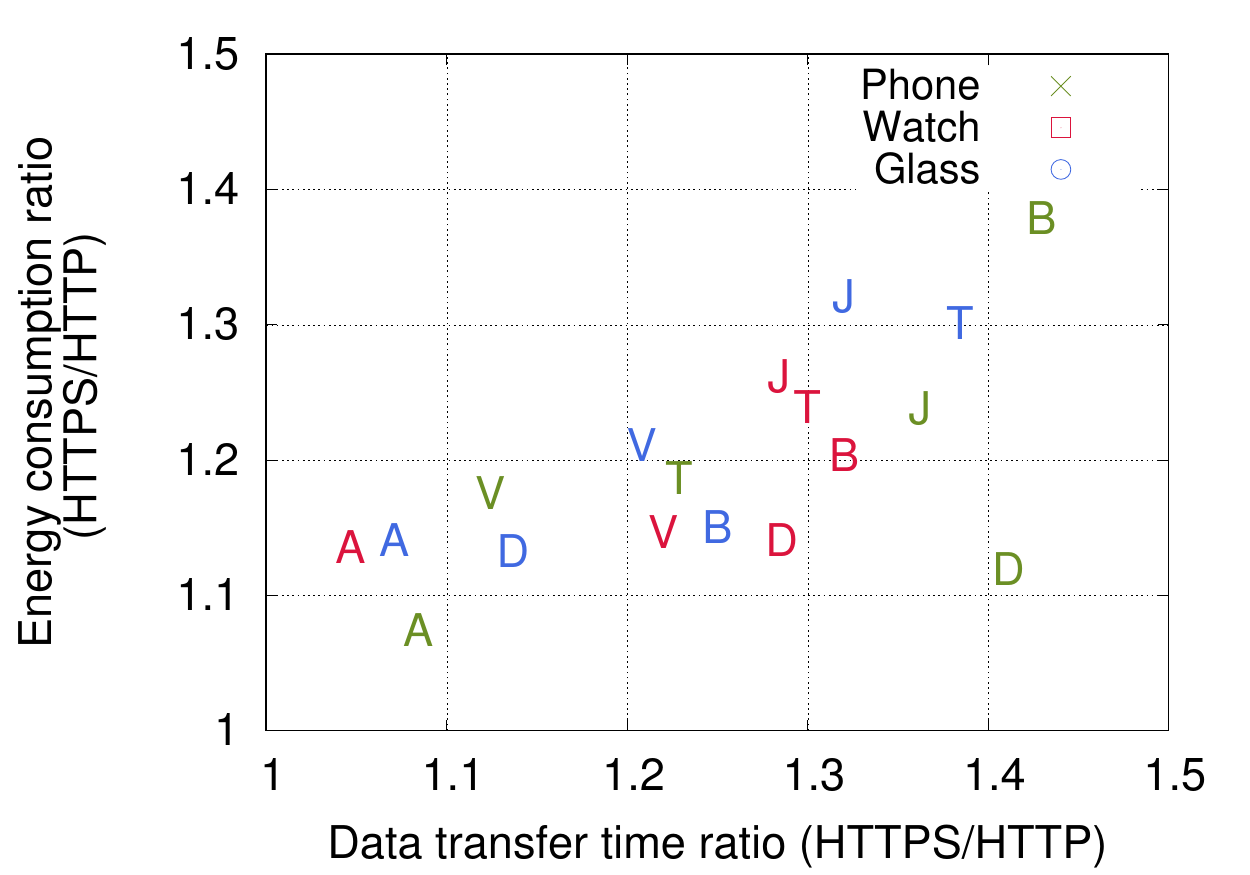}}\vspace{-2mm}
\caption{TLS cost distribution of websites in the three devices. The marker identifies a website (Table.~\ref{tab:websites}) and the device.}  \vspace{-4mm}
\label{fig:internetscatter} 
\end{figure*}

In this section, we present a set of experiments performed with Internet websites to understand the impact of TLS when multiple parallel TCP connections and multiple external servers are involved in a data exchange.
We first analyze the properties of data exchange when mobile devices interact with web content (i.e., number of connections and servers involved in the data exchange and RTT). We then investigate the role of the main data exchange parameters (i.e. number of servers, number of connections) on the TLS cost. For all experiments in this section, we follow the experimental methodology \textsf{Exp3}, using the websites reported in Table~\ref{tab:websites}.

\subsection{Understanding Data Exchange with Web Servers} 

We start our analysis by investigating the main characteristics of the data exchange between the devices and external websites. The goal is to identify the main parameters that may affect the cost of HTTPS and quantify their effect.

Figure~\ref{fig:bar}, ~\ref{fig:energybarc} and ~\ref{fig:barb} depicts the average number of connections and servers involved in each data exchange for HTTP and HTTPS for each device type.  
As expected, we observe the number of servers and connections involved in the communication vary among websites. Specifically, the numbers of servers ranges between 1 and 7, while the number of connections is between 4 to 13. Indeed, those parameters are specific to each website and are related to the number of embedded objects and the servers from which they are delivered to the users. 

We also notice that the same websites on different devices show a different number of connections and servers, even for the same protocol. For instance, \texttt{\url{https://www.apple.com}} (A) opens 6, 7, and 8 connections for HTTPS in watch, glass and phone respectively. This is due to two main factors. First, websites provide different versions to users according to device type. These versions may differ on the data exchange parameters (e.g., servers, connections), but can also differ in the actual content (e.g., embedded objects, main page). Second, different devices and browsers have a different level of support for websites' features (e.g., Javascript, Flash).

\subsection{Analysis of the Cost of TLS in the Wild}
Figures~\ref{fig:bar}, ~\ref{fig:energybarc} and ~\ref{fig:barb} show that the number of servers and connections varies between HTTP and HTTPS for most websites due to the reasons that were explained in the previous subsection. It is not possible however, to identify a clear trend, as the number of servers and connections increases with TLS for some websites while it decreases for others. As an explanation for this behaviour, we note that the number of connections basically depends on TCP connection reusability, where if there is an existing opened connection, the browser tries to use that connection to fetch a new web object rather opening a new connection. Most websites have different data exchange parameters for HTTP and HTTPS, and that those parameters are specific to each website.

We report the average Round Trip Time (RTT) ratio ($\frac{\text{HTTPS}}{\text{HTTP}}$) for each website in Figure~\ref{fig:rtt}. The RTT value always increases with HTTPS (i.e., from $\sim$10\% to $\sim$90\%), due to the additional processing required overhead at both servers and clients for data encryption/decryption. Intuitively, we expect low resource wearable devices to have a larger increase in RTT, which is not
what we observe, as the increase is approximately similar for all devices. This suggests that low resource  client devices have a lower impact on the RTT than the network and server side.

Figure~\ref{fig:scatsize} depicts the ratio ($\frac{\text{HTTPS}}{\text{HTTP}}$) of \textit{downloaded bytes} for each website and device. The larger the size of the website, the lower the cost of TLS in terms of \textit{downloaded bytes} (i.e., the ratio is $\sim$1.5 when the file size is 100KB and ratio goes down to $\sim$1.05 when file size increases up to 1MB). This confirms that the trend noticed in the \textsf{Exp1} also holds for data exchange with Internet websites. Among the measured websites, \texttt{\url{www.terraclicks.com}} (T) has the smallest in size,  yet TLS has greatest performance impact. Moreover, as highlighted in the previous section, since websites provide different content for different devices and hence use a different number of connections in those devices, the volume of data exchanged is different. Therefore the \textit{downloaded bytes} ratio ($\frac{\text{HTTPS}}{\text{HTTP}}$) varies across devices for the same website.

In Figure~\ref{fig:scattime}, we show the relation between the number of connections and the \textit{downloaded bytes} ratio ($\frac{\text{HTTPS}}{\text{HTTP}}$) for each website across devices. As expected, due to extra bytes added by the TLS handshake for each and every connection in HTTPS, the $\frac{\text{HTTPS}}{\text{HTTP}}$ ratio increases with the number of connections (i.e., $\sim$1.05 for 4 connections to $\sim$1.4 for 12 connections). 
However, in cases such as \texttt{\url{www.terraclicks.com}} (T), the byte overhead is dominated by the small size of the website rather than the number of connections. Due to the different sizes of certificates and different configurations of TLS parameters supported by different servers, the number of extra bytes added to each connection varies. This causes slight deviations to the expected trend for the ratio ($\frac{\text{HTTPS}}{\text{HTTP}}$) vs number of connections.

As observed in \textsf{Exp1}, the \textit{data exchange time} ratio impacts the \textit{energy consumption} ratio the most. Therefore, we analyze in Figure~\ref{fig:scatenergy} the \textit{energy consumption} ratio as a linear function of the \textit{data exchange time} ratio (in the range $\sim$1.05 to $\sim$1.5). As highlighted in \textsf{Exp1}, TLS handshake time and lower data transferring rate in HTTPS are the major causes of the additional duration in HTTPS.

\section{Conclusion and Main findings}
\label{sec:summary}

We ran our experiments of downloading files over HTTP and HTTPS from an internal server which we have the control over and also downloading 6 of the most popular websites according to Alexa ranking that support both HTTP and HTTPS. We consider the amount of  \textit{downloaded data}, the \textit{data transfer time} and the \textit{energy consumption} as the main metrics for our analysis. 

The main finding of the paper is that
\emph{secure direct Internet connectivity via WiFi on wearables is feasible, improves performance and reduces energy footprint.} In light of that, we provide the following recommendations to wearable application/system developers:

\begin{itemize}

\item Although wearables have the ability to connect to the Internet directly, most of the current wearable applications leverage the smartphone to connect to Internet. However, this requires the transfer of data from wearables to the smartphone via Bluetooth, and results increase in both  \textit{data transfer time} and \textit{energy consumption}. 
 
By using direct secure Internet communication instead of Bluetooth relaying for a typical website download, it is possible to achieve $\sim$78\% energy savings on the wearable (as well as 100\% in the smartphone) and reduce elapsed time by $\sim$40\%.
\emph{Therefore, we recommend app developers utilize direct Internet connectivity via WiFi from wearables eliminating the additional costly step of Bluetooth relaying.}

\item Despite the resource constraints on wearable devices, the relative cost of HTTPS, e.g., the ratio $\frac{\text{HTTPS}}{\text{HTTP}}$, in terms of \textit{data transfer time} is comparable on wearables and smartphones for all the sizes. The amount of \textit{downloaded data} is the same for any size of data download, if all the TLS algorithms used are the same for the devices.
Thus, standalone secure wearable applications can be practically realized utilizing the existing secure protocols such as HTTPS.

\end{itemize}

Moreover, we have the following observations that hold true for both smartphones and wearables.
\begin{itemize}

\item As expected, the impact of HTTPS is mostly driven by the size of the object (e.g., webpage, file) and number of TCP connections, and it is dominated by the TLS handshake processing and delays.
For small objects, TLS induces a significant increase in the \textit{energy consumption}, the \textit{data transfer time}, and the volume of \textit{downloaded data}. However, for large objects, the TLS cost is relatively smaller, and can be in the order of 15\%. These costs also increase with number of connections and servers.
The TLS handshake is the main contributor to the additional cost of HTTPS, while data encryption/decryption has a much lower impact. Therefore, the traditional approaches used in the mobile communications such as bulk transferring and reducing the number of TLS sessions can be exploited to increase the performance. 

\item The choice of TLS algorithms has a limited impact on the overall TLS cost. Key exchange is the stage that can be influenced the most by the choice of the algorithm and therein lies a security/performance trade-off. Conversely, different bulk cipher algorithms provide similar performance, and the one guaranteeing stronger security can thus be selected without adding much overhead.

\end{itemize}

Finally, important directions for future work are to study the cost of HTTPS in the cellular interface when it becomes common in consumer wearables, and to evaluate the performance of  other commonly used protocols  such as HTTP/2 and UDP on wearables.

\balance

\bibliographystyle{acm-sigchi}
\bibliography{references}
\end{document}